\RequirePackage{fix-cm}
\RequirePackage{multirow}
\documentclass[smallextended]{svjour3}    
\PassOptionsToPackage{numbers}{natbib}
\usepackage{natbib}
\usepackage{url}
\usepackage{graphicx}
\usepackage{array}
\usepackage{multirow}
\usepackage{verbatim}
\usepackage{hyphenat}
\usepackage{float}
\usepackage{tcolorbox}
\usepackage{enumitem}
\usepackage{balance}
\usepackage{todonotes}
\usepackage{threeparttable}
\usepackage{listings}
\usepackage{booktabs}
\usepackage{appendix}

\usepackage{todonotes}
\usepackage{footmisc}

\definecolor{gray50}{gray}{.5}
\definecolor{gray40}{gray}{.6}
\definecolor{gray30}{gray}{.7} 
\definecolor{gray20}{gray}{.8}
\definecolor{gray10}{gray}{.9}
\definecolor{gray05}{gray}{.95}

\newlength\Linewidth
\def\findlength{\setlength\Linewidth\linewidth
	\addtolength\Linewidth{-4\fboxrule}
	\addtolength\Linewidth{-3\fboxsep}
}
\newenvironment{rqbox}{\par\begingroup
	\setlength{\fboxsep}{5pt}\findlength
	\setbox0=\vbox\bgroup\noindent
	\hsize=0.95\linewidth
	\begin{minipage}{0.95\linewidth}\normalsize}
	{\end{minipage}\egroup
	\textcolor{gray20}{\fboxsep1.5pt\fbox
		{\fboxsep5pt\colorbox{gray10}{\normalcolor\box0}}}
	\endgroup\par\noindent
	\normalcolor\ignorespacesafterend}

\newcommand{\rb}[1]{
	
	\begin{tcolorbox}[colback=gray!05,
		colframe=black,
		width=\columnwidth,
		arc=3mm, auto outer arc,
		boxrule=0.5pt,
		]
		#1
	\end{tcolorbox}
}

\newcounter{Finding}
\stepcounter{Finding}

\newcommand{\roundedbox}[1]{
	\rb{
		\noindent
		\textit{\textbf{Finding \theFinding}. #1}
	}
	\stepcounter{Finding}
}

\begin{document}
\title{A Critical Comparison on Six Static Analysis Tools: Detection, Agreement, and Precision}
\titlerunning{A Critical Comparison on Six Static Analysis Tools}

                           
\author{Valentina Lenarduzzi \and Savanna Lujan \and Nyyti Saarim\"{a}ki \and Fabio Palomba}

\institute{Valentina Lenarduzzi \at
              LUT University, Finland \\
              \email{valentina.lenarduzzi@lut.fi}    
           \and
           Savanna Lujan and Nyyti Saarim\"{a}ki  \at
           Tampere University, Finland \\
           \email{savanna.lujan@tuni.fi} \\
            \email{nyyti.saarimaki@tuni.fi} 
           \and
           Fabio Palomba \at
           SeSa Lab - University of Salerno, Italy \\
           \email{fpalomba@unisa.it}
}

\date{Received: date / Accepted: date}

\maketitle

\begin{abstract}
\textit{Background.} Developers use Automated Static Analysis Tools (ASATs) to control for potential quality issues in source code, including defects and technical debt. Tool vendors have devised quite a number of tools, which makes it harder for practitioners to select the most suitable one for their needs. To better support developers, researchers have been conducting several studies on ASATs to favor the understanding of their actual capabilities. 

\noindent\textit{Aims.} Despite the work done so far, there is still a lack of knowledge regarding (1) which source quality problems can actually be detected by static analysis tool warnings, (2) what is their agreement, and (3) what is the precision of their recommendations. We aim at bridging this gap by proposing a large-scale comparison of six  popular static analysis tools for Java projects: Better Code Hub, CheckStyle, Coverity Scan, Findbugs, PMD, and SonarQube. 

\noindent\textit{Method.} We analyze 47 Java projects and derive a taxonomy of warnings raised by 6 state-of-the-practice ASATs. To assess their agreement, we compared them by manually analyzing - at line-level - whether they identify the same issues. Finally, we manually evaluate the precision of the tools. 

\noindent\textit{Results.} The key results report a comprehensive taxonomy of ASATs warnings, show little to no agreement among the tools and a low degree of precision.

\noindent\textit{Conclusions.} We provide a taxonomy that can be useful to researchers, practitioners, and tool vendors to map the current capabilities of the tools. Furthermore, our study provides the first overview on the agreement among different tools as well as an extensive analysis of their precision.

\end{abstract}

\section{Introduction}
\label{Intro}

Automated Static Analysis Tools (ASATs) are instruments that analyze characteristics of the source code without executing it, so that they can discover potential source code quality issues \cite{Ernst2015}. These tools are getting more popular as they are becoming easier to use---especially in continuous integration pipelines~\cite{zampetti2017open}---and there is a wide range to choose from~\cite{vassallo2019developers}. However, as the number of available tools grows, it becomes harder for practitioners to choose the tool (or combination thereof) that is most suitable for their needs~\cite{thomas2016questions}. 

To help practitioners with this selection process, researchers have been conducting empirical studies to compare the capabilities of existing ASATs \cite{mantere2009comparison,wilander2002comparison}. Most of these investigations have focused on (1) the features provided by the tools, e.g., which maintainability dimensions can be tracked by current ASATs, (2) comparing specific aspects considered by the tools, such as security \cite{antunes2009comparing,mclean2012comparing} or concurrency defects \cite{al2010comparing}, and (3) assessing the number of false positives given by the available static analysis tools \cite{johnson2013don}.

Recognizing the effort spent by the research community, which led to notable advances in the way tool vendors develop ASATs, we herein notice that our knowledge on the capabilities of the existing SATs is still limited. More specifically, in the context of our research we point our that three specific aspects are under-investigated: (1) which source quality problems can actually be detected by static analysis tools, (2) what is the agreement among different tools with respect to source code marked as potentially problematic, and (3) what is the precision with which a large variety of the available tools provide recommendations. An improved knowledge of these aspects would not only allow practitioners to take informed decisions when selecting the tool(s) to use, but also researchers/tool vendors to enhance them and improve the level of support provided to developers.

In this paper, we aim to address this gap of knowledge by designing and conducting a large-scale empirical investigation into the detection capabilities of six popular state-of-the-practice ASATs, namely SonarQube, Better Code Hub, Coverity Scan, Findbugs, PMD, and CheckStyle.\footnote{ASATs verify code compliance with a specific set of warnings that, if violated, can introduce an issue in the code. This issue can be accounted for as ``source code quality issue'': as such, in the remaining the paper we use this term when referring to the output of the considered tools.} Specifically, we run the considered tools against a corpus of 47 projects from the Qualitas Corpus dataset and (1) depict which source quality problems can actually be detected by the tools, (2) compute the agreement among the recommendations given by them at line-level, and (3) manually compute the precision of the tools. 

The key results of the study report a taxonomy of static analysis tool warnings, but also show that, among the considered tools, SonarQube is the one able to detect most of the quality issues that can be detected by the other ASATs. However, when considering the specific quality issues detected, there is little to no agreement among the tools, indicating that different tools are able to identify different forms of quality problems. Finally, the precision of the considered tools ranges between 18\% and 86\%, meaning that the practical usefulness of some tools is seriously threatened by the presence of false positives---this result corroborates and enlarges previous findings \cite{johnson2013don} on a larger scale and considering a broader set of tools.

Based on our findings, our paper finally discusses and distills a number of lessons learned, limitations, and open challenges that should be considered and/or addressed by both the research community and tool vendors. 

\smallskip
\noindent \textbf{Structure of the paper.} The study setting is described in Section~\ref{Case Study}. The results are presented in Section~\ref{Results} and discussed in Section~\ref{Discussion}. Section~\ref{Threats} identifies the threats to the validity of our study, while Section~\ref{RW} presents related works on static analysis tools. Finally, in Section~\ref{Conclusion} we draw conclusions and provide an outlook on our future research agenda.

\section{Empirical Study Design}
\label{Case Study}
We designed our empirical study as a case study based on the guidelines defined by Runeson and H\"{o}st~\cite{Runeson2009}. The following sections describe the goals and specific research questions driving our empirical study as well as the data collection and analysis procedures. 

\subsection{Goal and Research Questions}
\label{Goal and Research Questions}
The \emph{goal} of our empirical study is to compare state-of-the-practice ASATs \emph{with the aim of} assessing their capabilities when detecting source code quality issues with respect to (1) the types of problems they can actually identify; (2) the agreement among them, and (3) their precision. Our ultimate \emph{purpose} is to enlarge the knowledge available on the identification of source code quality issues with ASATs from the \emph{perspective} of both researchers and tool vendors. The former are interested in identifying areas where the state-of-the-art tools can be improved, thus setting up future research challenges, while the latter are instead concerned with assessing their current capabilities and possibly the limitations that should be addressed in the future to better support developers. 

More specifically, our goal can be structured around three main research questions (\textbf{RQ$_s$}). As a first step, we aimed at understanding what kind of issues different tool warnings detect when run on source code. An improved analysis of this aspect may extend our knowledge on whether and how various types of quality issues are identified by the existing tools. Hence, we asked:

\begin{center}	
	\begin{rqbox}
		\textbf{RQ$_1$.} \emph{What source code quality issues can be detected by Automated Static Analysis Tools?}
	\end{rqbox}	 
\end{center}

Once we had characterized the tools with respect to what they are able to identify, we proceeded with a finer-grained investigation aimed at measuring the extent to which ASATs agree with each other. Regarding this aspect, further investigation would not only benefit tool vendors who want to better understand the capabilities of their tools compared to others, but would also benefit practitioners who would like to know whether it is worth using multiple tools within their code base. Moreover, we were interested in how the issues from different tools overlap with each other. We wanted to determine the type and number of overlapping issues, but also whether the overlapping is between all tools or just a subset.

\begin{center}	
	\begin{rqbox}
		\textbf{RQ$_2$.} \emph{What is the agreement among different Automated Static Analysis Tools when detecting source code quality issues?}
	\end{rqbox}	 
\end{center}

Finally, we focused on investigating the potential usefulness of the tools in practice. While they could output numerous warnings that alert developers of the presence of potential quality problems, it is still possible that some of these warnings might represent false positive instances, i.e., that they wrongly recommend source code entities to be refactored/investigated. Previous studies have highlighted the presence of false positives as one of the main problems of the tools currently available~\cite{johnson2013don}; our study aims at corroborating and extending the available findings, as further remarked in Section \ref{DataAnalysis}. 
\begin{center}	
	\begin{rqbox}
		\textbf{RQ$_3$.} \emph{What is the precision of Static Analysis Tools?}
	\end{rqbox}	 
\end{center}

All in all, our goal was to provide an updated view on this matter and understand whether, and to what extent, this problem has been mitigated in recent years or whether there is still room for improvement. 



\subsection{Context of the Study}
\label{Context}
The \emph{context} of our study consisted of software systems and tools. In the following, we describe our selection.

\subsubsection*{Project Selection}
We selected projects from the Qualitas Corpus collection of software systems (Release 20130901), using the compiled version of the Qualitas Corpus~\cite{Terra2013}. 

The dataset contains 112 Java systems with 754 versions, more than 18 million LOCs, 16,000 packages, and 200,000 classes analyzed. 
Moreover, the dataset includes projects from different contexts such as IDEs, databases, and programming language compilers. More information is available in~\cite{Terra2013}. In our study, we considered the ``r'' release of each of the 112 available systems. 
Since two of the automated static analysis tools considered, i.e., Coverity Scan and Better Code Hub, require permissions in the GitHub project or the upload of a configuration file, we privately uploaded all 112 projects to our GitHub account in order to enable the analysis\footnote{The GitHub projects, with the related configuration adopted for executing the tools, will be made public in the case of acceptance of this paper.}. 







\subsubsection*{ASATs Selection} We selected the six tools described below. The choice of focusing on those specific tools was driven by the familiarity of the authors with them: this allowed us to (1) use/run them better (e.g., by running them without errors) and (2) analyze their results better, for instance by providing qualitative insights able to explain the reasons behind the achieved results. The analysis of other tools is already part of our future research agenda. 

\smallskip
\textbf{SonarQube}\footnote{\label{Sonar}http://www.sonarsource.org/}
\label{SonarQube}
is one of the most popular open-source static code analysis tools for measuring code quality issues. It is provided as a service by the \texttt{sonarcloud.io} platform or it can be downloaded and executed on a private server.
SonarQube computes several metrics such as number of lines of code and code complexity, and verifies code compliance with a specific set of ``coding warnings'' defined for most common development languages. If the analyzed source code violates a coding warning, the tool reports an ``issue''. The time needed to remove these issues is called remediation effort. 

SonarQube includes reliability, maintainability, and security warnings. 
Reliability warnings, also named \textit{Bugs}, create quality issues that ``represent something wrong in the code'' and that will soon be reflected in a bug.  \textit{Code smells} are considered  ``maintainability-related issues'' in the code that decrease code readability and code modifiability. It is important to note that the term ``code smells'' adopted in SonarQube does not refer to the commonly known term code smells defined by Fowler et al.~\cite{Fowler1999}, but to a different set of warnings. 


\smallskip
\textbf{Coverity Scan}\footnote{\url{https://scan.coverity.com/}}
\label{CoverityScan} 
is another common open-source static analysis tool. The code build is analyzed by submitting the build to the server through the public API.
The tool detects defects and vulnerabilities that are grouped by \textit{categories} such as: resource leaks, dereferences of NULL pointers, incorrect usage of APIs, use of uninitialized data, memory corruptions, buffer overruns, control flow issues, error handling issues, incorrect expressions, concurrency issues, insecure data handling, unsafe use of signed values, and use of resources that have been freed\footnote{\url{https://scan.coverity.com/faq\#what-is-coverity-scan}}. For each of these \textit{categories}, there are various issue \textit{types} that explain more details about the defect. In addition to issue \textit{types}, issues are grouped based on \textit{impact}: low, medium, and high.
The static analysis applied by Coverity Scan is based on the examination of the source code by determining all possible paths the program may take. This gives a better understanding of the control and data flow of the code\footnote{\url{https://devguide.python.org/coverity}}.

\smallskip
\textbf{Better Code Hub}\footnote{\url{https://bettercodehub.com/}} is also a commonly used static analysis tool that assesses code quality. The analysis is done through the website's API, which analyzes the repository from GitHub. The default configuration file can be modified for customization purposes. Code quality is generally measured based on structure, organization, modifiability, and comprehensibility. 

This is done by assessing the code against ten \textit{guidelines}: write short units of code, write simple units of code, write code once, keep unit interfaces small, separate concern in modules, couple architecture components loosely, keep architecture components balanced, keep your code base small, automate tests, and write clean code. Out of the ten \textit{guidelines}, eight \textit{guidelines} are grouped based on type of \textit{severity}: medium, high, and very high. \textit{Compliance} is rated on a scale from 1-10 based on the results\footnote{\url{https://pybit.es/bettercodehub.html}}.

Better Code Hub static analysis is based on the analysis of the source code against heuristics and commonly adopted coding conventions. This gives a holistic view of the health of the code from a macroscopic perspective. 

\label{BetterCodeHub}

\smallskip
\textbf{Checkstyle}\footnote{\url{https://checkstyle.org}}
\label{Checkstyle} is an open-source developer tool that evaluates Java code quality. The analysis is done either by using it as a side feature in Ant or as a command line tool. Checkstyle assesses code according to a certain coding standard, which is configured according to a set of \textit{checks}. Checkstyle has two sets of style configurations for standard \textit{checks}: Google Java Style\footnote{\url{https://checkstyle.sourceforge.io/google_style.html}} and Sun Java Style\footnote{\url{https://checkstyle.sourceforge.io/sun_style.html}}. In addition to standard \textit{checks} provided by Checkstyle, customized configuration files are also possible according to user preference.~\footnote{\url{https://checkstyle.sourceforge.io/index.html}}

These \textit{checks} are classified under 14 different categories: annotations, block checks, class design, coding, headers, imports, javadoc comments, metrics, miscellaneous, modifiers, naming conventions, regexp, size violations, and whitespace. Moreover, the violation of the \textit{checks} are grouped under two severity levels: error and warning\footnote{\url{https://checkstyle.sourceforge.io/checks.html}}, with the first reporting actual problems and the second possible issues to be verified.

\smallskip
\textbf{FindBugs}\footnote{\url{http://findbugs.sourceforge.net}}
\label{FindBugs} is a static analysis tool for evaluating Java code, more precisely Java bytecode. The analysis is done using the GUI, which is engaged through the command line. The analysis applied by the tool is based on detecting \textit{bug patterns}. According to FindBugs, the \textit{bug patterns} arise for the following main reasons: difficult language features, misunderstood API features, misunderstood invariants when code is modified during maintenance, and garden variety mistakes.\footnote{\url{http://findbugs.sourceforge.net/findbugs2.html}} \footnote{\url{http://findbugs.sourceforge.net/factSheet.html}}
Such \textit{bug patterns} are classified under 9 different categories: bad practice, correctness, experimental, internationalization, malicious code vulnerability, multithreaded correctness, performance, security, and dodgy code. Moreover, the \textit{bug patterns} are ranked from 1-20. Rank 1-4 is the \textit{scariest} group, rank 5-9 is the \textit{scary} group, rank 10-14 is the \textit{troubling} group, and rank 15-20 is the \textit{concern} group\footnote{\url{http://findbugs.sourceforge.net/bugDescriptions.html}}.

\smallskip
\textbf{PMD}\footnote{\url{https://pmd.github.io/latest/}}
\label{FindBugs} is a static analysis tool mainly used to evaluate Java and Apex, even though it can also be applied to six other programming languages. The analysis is done through the command line using the binary distributions. PMD uses a set of \textit{warnings} to assess code quality according to the main focus areas: unused variables, empty catch blocks, unnecessary object creation, and more. There are a total of 33 different warning set configurations\footnote{\url{https://github.com/pmd/pmd/tree/master/pmd-java/src/main/resources/rulesets/java}} for Java projects. The warning sets can also be customized according to the user preference\footnote{\url{https://pmd.github.io/latest/index.html}}. These \textit{warnings} are classified under 8 different categories: best practices, code style, design, documentation, error prone, multi threading, performance, and security. Moreover, the violations of  \textit{warnings} are measured on a priority scale from 1-5, with 1 being the most severe and 5 being the least\footnote{\url{https://pmd.github.io/latest/pmd$_$rules$_$java.html}}.

\subsection{Study Setup and Data Collection}
\label{StudySetup}
This section describes the study setup used to collect the data from each tool and the data collection process. We analyzed a single snapshot of each project, considering the release available in the dataset for each of 112 systems.

\smallskip
\textbf{SonarQube}. We first installed  SonarQube LTS 6.7.7 on a private server having 128 GB RAM and 4 processors. However, because of the limitations of the open-source version of SonarQube, we are allowed to use only one core, therefore more cores would have not been beneficial for our scope. We decided to adopt the LTS version (Long-Time Support) since this is the most stable and best-supported version of the tool. 

We  executed SonarQube on each project using SDK 8 (Oracle) and the \texttt{sonar-scanner} package version 4.2. Each project was analyzed using the original sources and the binaries provided in the dataset.
Moreover, we configured the analysis (in the file \texttt{sonar-project.properties}) 
reporting information regarding the project key, project name, project version, source directories, test directories, binary directories, and library directories. 
It is important to note that the analysis performed using the original binaries reduced the issues of compilation errors and missing libraries. Moreover, it also helped to reduce issues related to false positives\footnote{\url{https://docs.sonarqube.org/latest/analysis/languages/java/}}.
Once the configuration file was set up, the analysis began by running sonar scanner in the root directory of the project: 

\texttt{\$ sonar-scanner}.

After all the projects had been analyzed, we extracted the data related to the projects using the \textit{''SonarQube Exporter''} tool\footnote{\url{https://github.com/pervcomp/SonarQube-Exporter/releases}}. This makes it possible to extract SonarQube project measures and issues in CSV format. 
Under the \texttt{target} directory, the \texttt{extraction} directory was made. The \texttt{CSV} files were then extracted from the server by running the following command:

\smallskip
\noindent\texttt{\$java-jar csv-from-sonar-0.4.1-jar-with-dependencies.jar}


\smallskip
All projects starting with “QC:” were extracted from the server by clicking “Save issues” and “Save measures”. This exported \texttt{CSV} files into the \texttt{extraction} directory for each project.

\smallskip
\textbf{Coverity Scan}.
The projects were registered in Coverity Scan (version 2017.07) by linking the \texttt{github} account and adding all the projects to the profile. 
Coverity Scan was set up by downloading the tarball file from \url{https://scan.coverity.com/download} and adding the \texttt{bin} directory of the installation folder to the path in the \texttt{.bash\_profile}. Afterwards the building process began, which was dependent on the type of project in question. Coverity Scan requires to compile the sources with a special command. Therefore, we had to compile them, instead of using the original binaries. For our projects, the following commands were used in the directory of the project where the build.xml and pom.xml files reside:

\begin{itemize}
    \smallskip
    \item \texttt{ant} (build.xml) for building and deploying Java projects. The projects had various Java versions, so the appropriate Java version was installed according to the documentation (if available) for building. 
    
    \smallskip
    \begin{itemize}
        \item \texttt{\$ cov-build –dir cov-int ant jar }
        
        \smallskip
        \item \texttt{\$ cov-build –dir cov-int ant}
    \end{itemize}
    
    \smallskip
    \item \texttt{maven} (pom.xml) for building projects. The appropriate \texttt{maven} version was used according to the documentation.
    \begin{itemize}
        \smallskip
        \item \texttt{\$ cov-build –dir cov-int mvn install}
        
        \smallskip
        \item \texttt{\$ cov-int –dir cov-int mvn clean install}
    \end{itemize}
\end{itemize}

After building the project, if over 85\% of tests were successfully executed, a \texttt{tar} archive of the \texttt{cov-int} build folder was created by running:

\smallskip
\texttt{\$ tar czvf myproject.tgz cov-int}

\smallskip
Once the \texttt{tar} archive was created, the file was submitted through the project dashboard at \url{https://scan.coverity.com/projects/myproject} to ``Submit Build''. The analysis was then performed, and the results were displayed on the dashboard. The data regarding the analysis was extracted from the dashboard from ``View Defects'', where the detailed report regarding the analysis results was located. Under Outstanding Issues -$>$ Export CSV, the \texttt{CSV} file containing all issues of the project was downloaded for each project. 

\smallskip
\textbf{Better Code Hub.}
The \texttt{.bettercodehub.yml} files were configured by defining the \texttt{component\_depth}, \texttt{languages}, and \texttt{exclusions}. The \texttt{exclusions} were defined so that they would exclude all directories that were not source code, since Better Code Hub only analyzes source code. The analysis was conducted between January 2019 and February 2019, so the Better Code Hub version hosted during that time was used. 


Once the configuration file had been created, it was saved in the root directory of the project. The local changes to the project were added, committed, and pushed to GitHub. Afterwards, the project analysis started from Better Code Hub’s dashboard, which is connected to the GitHub account \url{https://bettercodehub.com/repositories}. The analysis started by clicking the ``Analyze'' button on the webpage. The results were then shown on the webpage. Once all the projects had been analyzed, the data was sent in one CSV file. All projects containing more than 100,000 lines of code were not analyzed, due to Better Code Hub’s limit.

\smallskip
\textbf{Checkstyle}.
The \texttt{JAR} file for the Checkstyle analysis was downloaded directly from Checkstyle's website\footnote{\label{CheckstyleDownload}\url{https://checkstyle.org/#Download}} in order to engage the analysis from the command line. The executable \texttt{JAR} file used in this case was \texttt{checkstyle$>$8.30-all.jar}. In addition to downloading the \texttt{JAR} executable, Checkstlye offers two different types of warning sets for the analysis\footref{CheckstyleDownload}. 

For each of the warning sets, the configuration file was downloaded directly from Checkstyle's website\footnote{\url{https://github.com/checkstyle/checkstyle/tree/master/src/main/resources}}. In order to start the analysis, the files \texttt{checkstyle-8.30-all.jar} and the configuration file in question were saved in the directory where all the cloned repositories from Java Qualitas Corpus resided. To make the analysis more swift, a bash script was made to execute the analysis for each project in one go. This can be seen in Listing \ref{lst:list1}. 

\begin{lstlisting}[language=bash,caption={Checkstyle bash script tailored towards each warning set.}, label={lst:list1}]
#!/bin/bash
while read in; do java -jar checkstyle-8.30-all.jar 
-c /RULESET -f xml "$in"/ > "$in"_CS_RULESET.xml ; 
done < projectList.txt"
\end{lstlisting}

where \texttt{RULESET} represents the warning set used for the analysis, \texttt{"$\$$in"} represents the project name which is imported from projectList.txt, and \texttt{"$\$$in"\_CS\_RULESET.xml} represents the export file name of the analysis results in XML format. The text file projectList.txt consists of all the project names, in order to execute the analysis for all projects in one go. 
An example of how the projects were analyzed with Checkstyle according to the warning set Google Java Style\footnote{\url{https://github.com/checkstyle/checkstyle/blob/master/src/main/resources/google_checks.xml}} is show in Listing \ref{lst:list2}. 

\begin{lstlisting}[language=bash,caption={Example of Checkstyle bash script for Google Java Style configuration.}, label={lst:list2}]
#!/bin/bash
while read in; do java -jar checkstyle-8.30-all.jar 
-c /google_checks.xml -f xml "$in"/ > 
"$in"_CS_Google_Checks.xml ; 
done < projectList.txt"
\end{lstlisting}

\smallskip
\textbf{Findbugs}. FindBugs 3.0.1 was installed by running \texttt{brew install findbugs} in the command line. Once installed, the GUI was then engaged by writing \texttt{spotbugs}. From the GUI, the analysis was executed through \(\texttt{File} \rightarrow \texttt{New\hspace{0.125cm}Project}\). The classpath for the analysis was identified to be the location of the project directory. 

Moreover, the source directories were identified to be the project \texttt{JAR} executables. Once the classpath and source directories were identified, the analysis was engaged by clicking \texttt{Analyze} in the GUI. Once the analysis finished, the results were saved through \texttt{File} $\rightarrow$ \texttt{Save as} using the \texttt{XML} file format. 


\smallskip
\textbf{PMD}. PMD 6.23.0 was downloaded from GitHub\footnote{\url{https://github.com/pmd/pmd/releases/download/pmd$_$releases\%2F6.23.0/pmd-bin-6.23.0.zip}} as a zip file. After unzipping, the analysis was engaged by identifying several parameters: project directory, export file format, warning set, and export file name. In addition to downloading the zip file, PMD offers 32 different types of warning sets for Java written projects\footnote{\url{https://github.com/pmd/pmd/tree/master/pmd-java/src/main/resources/rulesets/java}}. We developed a bash script, shown in Listing \ref{lst:list3}, to engage the analysis for each project in one go. 

\begin{lstlisting}[language=bash,caption={PMD bash script tailored towards each warning set.}, label={lst:list3}]
#!/bin/bash
while read in; 
do $HOME/pmd-bin-6.23.0/bin/run.sh pmd -dir "$in"/
-f xml -R rulesets/java/RULESET  
-reportfile "$in"_PMD_RULESET.xml; 
done < projectList.txt"
\end{lstlisting}

The parameter HOME represents the full path where the binary resides, \texttt{"\$in"} represents the project name which is imported from projectList.txt, \texttt{RULESET} represents the warning set used for the analysis, and \texttt{"\$in"$\_$PMD$\_$RULESET.xml} represents the export file name of the analysis results in XML format. Just like in Listing \ref{lst:list1}, projectList.txt consists of all the project names. An example of how the projects were analyzed for the warning set Clone Implementation\footnote{\url{https://github.com/pmd/pmd/blob/master/pmd-java/src/main/resources/rulesets/java/clone.xml}} is show in Listing \ref{lst:list4}.

\begin{lstlisting}[language=bash,caption={Example of PMD bash script for Clone Implementation configuration.}, label={lst:list4}]
#!/bin/bash
while read in; 
do $HOME/pmd-bin-6.23.0/bin/run.sh pmd -dir "$in"/
-f xml -R rulesets/java/clone.xml  
-reportfile "$in"_PMD_Clone.xml; 
done < projectList.txt"
\end{lstlisting}

\subsection{Data Analysis}
\label{DataAnalysis}
In this section, we describe the analysis methods employed to address our research questions (\textbf{RQs}). 

\smallskip
\textbf{Source code quality issues identified by the tools (RQ$_1$).}
In order to determine the tool detection capabilities and warnings overlaps, we first identified warnings that can be detected by each tool, also considering type (if available), category, and severity. Then, we calculated how many warnings are violated in our projects. 

\smallskip
\textbf{Agreement among the tools (RQ$_2$).}
We expected that similar warnings should be violated in the same class, and in particular in the same position of the code. For the tools that provided information on the exact position (the lines of code where the warning is detected), we analyzed the agreement using the bottom-up approach. 
Therefore, we examined the overlapping positioning (start and end lines) of the warnings and identifying whether the warnings highlight the same kind of issue. Essentially pairing the warnings based on position and checking whether they have similar definitions.

In order to check whether the warning pairs identified in RQ$_1$ appear in the same position in the code, we scanned and checked in which classes each warning was violated. Then we counted how many times the warning pairs were violated in the same class. Only warning pairs identified across the tools were considered. warnings detected only by one tool were not considered. 

For example, there was a warning pair identified between SonarQube and Better Code Hub: ''SQ\_1 and BCH\_1''. Considering the warning pair detected by SonarQube (SQ) and Better Code Hub (BCH), we expected that these two warnings would be violated at least in the same class, since they identify the same kind of issue. So if for instance SQ\_1 is violated 100 times and BCH\_1 130 times, we would expect to find the warning pair violated at least 100 times in the same class. Hence, we can calculate the agreement between the warnings in the warning pair as follows:
\begin{equation}
    Agreement_r = \frac{\#SR}{\#issues_r} 
\end{equation}
where $\#SR$ is the number of instances in which a warning pair is detected together in the same class, and $\#issues_r$ is the overall number of issues of warning $r$ detected in the warning pair.

Agreement is calculated separately for both warnings in a warning pair. Perfect overlap is obtained if the agreement for both warnings is equal to one. This means that all the issues generated by those warnings are always detected in the same classes, and no issues are detected separately in a different class.  

After analyzing using the top-down approach from the definition level to the class level, we continued the analysis using the bottom-up approach from the line level  to the definition level. For each class, we considered the start and end lines of the issue and compared the degree to which the warnings overlap according to the lines affected.

As the granularity of the warnings varies between the tools, we checked what fraction of the affected lines are overlapping between the warnings, instead of requiring sufficient overlap between both warnings. This was done by selecting one issue at a time (reference) and comparing all other issues (comparison) to that. If the lines affected by the comparison warning were residing  within the lines affected by the reference warning, we defined the warnings as overlapped. To quantify the degree of overlapping, we used the percentage of the lines affected by the comparison issue that overlapped with the reference issue. The results were grouped based on four percentage thresholds: 100\%, 90\%, 80\% and 70\%.

\begin{figure}[H]
    \centering
    \includegraphics[width=0.7\columnwidth]{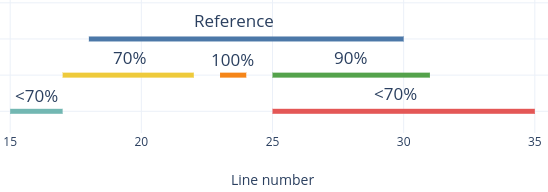}
    \caption{Determining overlapping warnings at ``line-level''.}
    \label{fig:thresholds}
    \vspace{-3mm}
\end{figure}

The concept is visualized in Figure~\ref{fig:thresholds}. The lines represent the issues in a code file, indicating the start and end of the affected code lines. The percentages represent the ratio of the lines affected by the warning that lie within the lines affected by the reference warning. Depending on the threshold used, different warnings are selected based on the overlapping percentage.

Unfortunately, only SonarQube, Better Code Hub, Checkstyle, PMD, and FindBugs provide information about the actual ''position'' of the detected warnings. They report information about the ''start  line'' and ''end line'' for each class. Regarding Coverity Scan, this information is only available in the web interface, but not in the APIs. Moreover, Coverity Scan licence does not allow to crawl the web interface. Since we have detected 8,828 Coverity Scan warnings violated in our projects, it would not have been visible to report this information manually. 


\smallskip
\textbf{Precision of the tools (RQ$_3$).} 
In our last research question, we aimed at assessing the precision of the considered tools. From a theoretical point of view, precision is defined as the ratio between the true positive source code quality issues identified by a tool and the total number of issues it detects, i.e., true positives plus false positive items (TPs + FPs). Formally, for each tool we computed precision as follows:

\begin{equation}
    precision = \frac{TPs}{TPs + FPs}
\end{equation}

It is worth remarking that our focus on precision is driven by recent findings in the field that showed that the presence of false positives is among the most critical barriers to the adoption of static analysis tools in practice \cite{johnson2013don,vassallo2019developers}. Hence, our analysis provides research community, practitioners, and tool vendors with indications on the actual precision of the currently available tools---and aims at possibly highlighting limitations that can be addressed by further studies. It is also important to remark that we do not assess recall, i.e., the number of true positive items identified over the total number of quality issues present in a software project, because of the lack of a comprehensive ground truth. We plan to create a similar dataset and perform such an additional evaluation as part of our future research agenda.

When assessing precision, a crucial detail is related to the computation of the set of true positive quality issues identified by each tool. In the context of our work, we conducted a manual analysis of the warnings highlighted by the six considered tools, thus marking each of them as true or false positive based on our analysis of (1) the quality issue identified and (2) the source code of the system where the issue was detected. Given the expensive amount of work required for a manual inspection, we could not consider all the warnings output by each tool, but rather focused on statistically significant samples. Specifically, we took into account a 95\% statistically significant stratified sample with a 5\% confidence interval of the 65,133, 8,828, 62,293, 402,409, 33,704, and 467,583 items given by Better Code Hub, Coverity Scan, SonarQube, Checkstyle, FindBugs, and PMD respectively: This step led to the selection of a set of 375 items from Better Code Hub, 367 from Coverity Scan, 384 from SonarQube, 384 from Checkstyle, 379 from FindBugs, and 380 from PMD.

To increase the reliability of this manual analysis, two of the authors of this paper (henceforth called the inspectors) first independently analyzed the warning samples. They were provided with a spreadsheet containing six columns: (1) the name of the static analysis tool the row refers to, i.e., Better Code Hub, Coverity Scan, Sonarqube, Checkstyle, FindBugs, and PMD; (2) the full path of the warning identified by the tool that the inspectors had to verify manually; and (3) the warning type and specification, e.g., the code smell. The inspectors' task was to go over each of the warnings and add a seventh column in the spreadsheet that indicated whether the warning was a true or a false positive. After this analysis, the two inspectors had a four-hour meeting where they discussed their work and resolved any disagreements: All the items marked as true or false positive by both inspectors were considered as actual true or false positives; in the case of a disagreement, the inspectors re-analyzed the warning in order to provide a common assessment. Overall, after the first phase of inspection, the inspectors reached an agreement of 0.84---which we computed using Krippendorff’s alpha Kr$_\alpha$ \cite{krippendorff2018content} and which is higher than 0.80, which has been reported to be the standard reference score for Kr$_\alpha$ \cite{antoine2014weighted}.

In Section \ref{results:rq3}, we report the precision values obtained for each of the considered tools and discuss some qualitative examples that emerged from the manual analysis of the sample dataset.

\subsection{Replicability} 
In order to allow the replication of our study, we have published the raw data in a replication package\footnote{\label{package} https://figshare.com/s/3c40dce067507b3e6c63}.


\section{Analysis of the Results}
\label{Results}
In this section, we report and discuss the results obtained when addressing our research questions (RQ$_s$). 

\smallskip
\noindent\textbf{RQ$_1$. What quality issues can be detected by  Static Analysis Tools?} Here we analyzed how many warnings are actually output by the considered static analysis tools as well as the types of issues they are able to discover.

\smallskip
\noindent\textbf{Static Analysis Tools Detection Capability}. We report the detection capability of each tool in terms of how many warnings can be detected, and the classification of internal warnings (e.g., type and severity). Moreover, we report the diffusion of the  warning in the selected projects. 

\smallskip
\textbf{Better Code Hub} detects a total of 10 warnings, of which 8 are grouped based on type and severity. Better Code Hub categorizes the 8 warnings under 3 types: \textit{RefactoringFileCandidateWithLocationList}, \textit{RefactoringFileCandidate}, and \textit{RefactoringFileCandidateWithCategory}. Of these 8 warnings, one is of \textit{RefactoringFileCandidateWithLocationList} type, six are of \textit{RefactoringFileCandidate} type, and one is of \textit{RefactoringFileCandidateWithCategory} type. In addition to the types, Better Code Hub assigns three possible severities to the warnings: \textit{Medium}, \textit{High}, and \textit{Very High}. Of these eight warnings, four were classified as \textit{Medium} severity, four as \textit{High} severity, and eight as \textit{Very High} severity. Some of the warnings have more than one severity possibly assigned to them.

\smallskip
\textbf{Checkstyle} detects a total of 173 warnings which are grouped based on type and severity. Checkstyle categorizes the 173 warnings under 14 types: \textit{Annotations}, \textit{Block Checks}, \textit{Class Design}, \textit{Coding}, \textit{Headers}, \textit{Imports}, \textit{Javac Comments}, \textit{Metrics}, \textit{Miscellaneous}, \textit{Modifiers}, \textit{Naming Conventions}, \textit{Regexp}, \textit{Size Violations}, and \textit{Whitespace}. Of these 173 warnings, 8 are of \textit{Annotations} type, 6 are of \textit{Block Checks} type, 9 are of \textit{Class Design} type, 52 are of \textit{Coding} type, 1 is of \textit{Headers} type, 8 are of \textit{Imports} type, 19 are of \textit{Javac Comments} type, 6 are of \textit{Metrics} type, 16 are of \textit{Miscellaneous} type, 4 are of \textit{Modifiers} type, 16 are of \textit{Naming Conventions} type, 4 are of \textit{Regexp} type, 8 are of \textit{Size Violations} type, and 16 are of \textit{Whitespace} type. In addition to these types, Checkstyle groups these checks under four different severity levels: \textit{Error}, \textit{Ignore}, \textit{Info}, and \textit{Warning}. The distribution of the checks with respect to the severity levels is not provided in the documentation. 


\smallskip
\textbf{Coverity Scan}'s total scope of detectable warnings as well as the classification is not known, since its documentation requires being a client. However, within the scope of our results, Coverity Scan detected a total of 130 warnings. These warnings were classified under three severity levels: \textit{Low}, \textit{Medium}, and \textit{High}. Of these 130 warnings, 48 were classified as \textit{Low} severity, 87 as \textit{Medium} severity, and 12 as \textit{High} severity. Like Better Code Hub, some of Coverity Scan's warnings have more than one severity type assigned to them. 

\smallskip
\textbf{Findbugs} detects a total of 424 warnings grouped based on type and severity. It categorizes the 424 warnings under 9 types: \textit{Bad practice}, \textit{Correctness}, \textit{Experimental}, \textit{Internationalization}, \textit{Malicious code vulnerability}, \textit{Multithreaded correctness}, \textit{Performance}, \textit{Security}, and \textit{Dodgy code}. Of these 424 warnings, 88 are of \textit{Bad practice} type, 149 are of \textit{Correctness}, 3 are of \textit{Experimental}, 2 are of \textit{Internationalization}, 17 are of \textit{Malicious code vulnerability}, 46 are of \textit{Multithreaded correctness}, 29 are of \textit{Performance}, 11 are of \textit{Security}, and 79 are of \textit{Dodgy code}. In addition to these types, Findbugs ranks these 'bug patterns' from 1-20. Rank 1-4 is the \textit{scariest} group, rank 5-9 is the \textit{scary} group, rank 10-14 is the \textit{troubling} group, and rank 15-20 is the \textit{concern} group.


\smallskip
\textbf{PMD} detects a total of 305 warnings which are grouped based on type and severity. PMD categorizes the 305 warnings under 8 types: \textit{Best Practices}, \textit{Code Style}, \textit{Design}, \textit{Documentation}, \textit{Error Prone}, \textit{Multithreading}, \textit{Performance}, and \textit{Security}. Of these 305 warnings, 51 are of \textit{Best Practices}, 62 are of \textit{Code Style}, 46 are of \textit{Design}, 5 are of \textit{Documentation}, 98 are of \textit{Error Prone}, 11 are of \textit{Multithreading}, 30 are of \textit{Performance}, and 2 are of \textit{Security} type. In addition to the types, PMD categorizes the warnings according to five priority levels (from P1 ``Change absolutely required'' to P5 ``Change highly optional'').  Rule priority guidelines for default and custom-made warnings can be found in the PMD project documentation.\footnote{\url{https://pmd.github.io/latest/}}

\smallskip
\textbf{SonarQube} LTS 6.7.7 detects a total of 413 warnings which are grouped based on type and severity. SonarQube categorizes the 413 warnings under 3 types: \textit{Bugs}, \textit{Code Smells}, and \textit{Vulnerabilities}. Of these 413 warnings, 107 warnings are classified as \textit{Bugs}, 272 as \textit{Code Smells}, and 34 as \textit{Vulnerabilities}. In addition to the types, SonarQube groups the warnings under 5 severity typers: \textit{Blocker}, \textit{Critical}, \textit{Major}, \textit{Minor}, and \textit{Info}. Considering the assigned severity levels, SonarQube detects 36 \textit{Blocker}, 61 \textit{Critical}, 170 \textit{Major}, 141 \textit{Minor}, and 5 \textit{Info} warnings. Unlike Better Code Hub and Coverity Scan, SonarQube has only one severity and classification type assigned to each rule. 

\begin{table}[h]
\tiny
    \centering
        \caption{\textbf{RQ$_1$.} Rule diffusion across the 47 projects. }
    \label{tab:Diffusion}
    \resizebox{1\linewidth}{!}{%
    \begin{tabular}{c|c|c|c|c|c|c|c|c|c} \hline 
         \textbf{Project Name} & \textbf{\#Classes} & \textbf{\#Methods} &      \textbf{SQ} &    \textbf{BCH} & \textbf{Coverity} & \textbf{Checkstyle} &      \textbf{PMD} & \textbf{FindBugs} &     \textbf{Total} \\ \hline 
            AOI &      865 &     2568 &   10865 &    924 &      123 &     250201 &   108458 &     1979 &    372550 \\
    Collections &      646 &     2019 &    4545 &    584 &       25 &      85501 &    39893 &      185 &    130733 \\
           Colt &      627 &     1482 &    7452 &    560 &       62 &     172034 &    47843 &        4 &    227955 \\
          Columba &     1288 &     2941 &    7030 &    662 &       70 &     166062 &    49068 &     1345 &    224237 \\
       DisplayTag &      337 &      683 &     853 &    452 &       22 &      32033 &    10137 &       32 &     43529 \\
        Drawswf &     1031 &     1079 &    3493 &    559 &       65 &     368052 &    22264 &       69 &    394502 \\
        Emma &      509 &      962 &    4451 &    648 &       55 &      68838 &    16524 &      172 &     90688 \\
       Findbugs &     1396 &     4691 &   12496 &    600 &      134 &     320087 &    90309 &     1068 &    424694 \\
       Freecol &     1569 &     4857 &    5963 &    607 &      337 &     127363 &    79588 &      704 &    214562 \\
       Freemind &     1773 &     3460 &    5698 &    662 &      112 &     128590 &    50873 &     1536 &    187471 \\
   Ganttproject &     1093 &     2404 &   12349 &    642 &       64 &      71872 &    36689 &      898 &    122514 \\
         Hadoop &     3880 &    10701 &   24125 &    682 &      665 &     284315 &   228966 &     1547 &    540300 \\
         HSQLDB &     1284 &     5459 &   14139 &    620 &      178 &     192010 &   109625 &      182 &    316754 \\
         Htmlunit &     3767 &     9061 &    5176 &    924 &      141 &      92998 &    59807 &      467 &    159513 \\
        Informa &      260 &      644 &     992 &    594 &       56 &      11276 &     9364 &      217 &     22499 \\
        Jag &     1234 &     1926 &    6091 &    301 &       56 &      24643 &    19818 &      408 &     51317 \\
        James &     4138 &     2197 &    6091 &    656 &       82 &     336107 &    29253 &       25 &    372214 \\
        Jasperreports &     2380 &     4699 &   17575 &    702 &      226 &     643076 &    96000 &     1420 &    758999 \\
        Javacc &      269 &      689 &    3693 &    504 &       29 &      24936 &    17784 &       39 &     46985 \\
        JBoss &     7650 &    18239 &   42190 &    415 &       51 &    1084739 &   377357 &     1158 &   1505910 \\
        JEdit &     2410 &     4918 &   15464 &    630 &      134 &     434183 &    93605 &       74 &    544090 \\
        JExt &     2798 &     2804 &    7185 &    585 &      339 &     276503 &    42693 &      125 &    327430 \\
        JFreechart &     1152 &     3534 &    6708 &    660 &       88 &     154064 &    89284 &      849 &    251653 \\
        JGraph &      314 &     1350 &    2577 &    666 &      128 &      98119 &    22516 &       41 &    124047 \\
        JGgraphPad &      433 &      916 &    2550 &    599 &       10 &      62230 &    18777 &       75 &     84241 \\
        JGgraphT &      330 &      696 &     922 &    562 &       35 &      23808 &    11147 &       15 &     36489 \\
        JGroups &     1370 &     4029 &   14497 &    602 &      391 &     265886 &    89601 &     1560 &    372537 \\
         JMoney &      183 &      455 &     575 &    426 &       52 &      15377 &     5639 &      118 &     22187 \\
         Jpf &      143 &      443 &     522 &    558 &       21 &      14736 &     7054 &       20 &     22911 \\
        JRefactory &     4210 &     5132 &   18165 &    580 &      129 &     207911 &   116452 &     2633 &    345870 \\
        Log4J &      674 &     1028 &    2042 &    625 &      125 &      40206 &    15463 &      162 &     58623 \\
        Lucene &     4454 &    10332 &   11332 &    707 &       85 &     627683 &   233379 &      585 &    873771 \\
        Marauroa &      266 &      777 &    1228 &    547 &       33 &      53616 &    10681 &      148 &     66253 \\
        Maven &     1730 &     4455 &    3110 &    642 &      121 &     225017 &    46620 &     1242 &    276752 \\
        Megamek &     3225 &     8754 &   14974 &    600 &      321 &     346070 &   174680 &     3430 &    540075 \\
        Myfaces\_core &     1922 &     5097 &   22247 &    312 &      121 &     619072 &   174790 &      790 &    817332 \\
        Nekohtml &       82 &      269 &     623 &    460 &       13 &      12987 &     3979 &       56 &     18118 \\
        PMD &     1263 &     3116 &    8818 &    616 &       50 &     109519 &    47664 &      543 &    167210 \\
        POI &     2276 &     8648 &   19463 &    903 &      771 &     476488 &   162045 &      792 &    660462 \\
        Proguard &     1043 &     1815 &    3203 &    646 &       23 &     115466 &    37221 &        6 &    156565 \\
        Quilt &      394 &      638 &    1075 &    386 &       13 &      16840 &     7488 &      170 &     25972 \\
        Sablecc &      251 &      886 &    4385 &    520 &       25 &      30840 &    19756 &      101 &     55627 \\
        Struts &     2598 &     6719 &    8878 &    616 &       57 &     231912 &   106513 &      253 &    348229 \\
        Sunflow &      227 &      670 &    1549 &    478 &       46 &      32251 &    20937 &       63 &     55324 \\
        Trove &      421 &      477 &     454 &    216 &       78 &      15507 &     5430 &      416 &     22101 \\
        Weka &     2147 &     6286 &   32258 &    604 &     1437 &     365535 &   195774 &     4118 &    599726 \\
        Xalan &     2174 &     4758 &   18362 &    844 &      232 &     330254 &   121685 &     1864 &    473241 \\ \hline 
                \textbf{Total} &    \textbf{74,486} &   \textbf{169,763} &  \textbf{418,433} &  \textbf{27,888} &     \textbf{7,431} &    \textbf{9,686,813} &  \textbf{3,380,493} &    \textbf{33,704} & \textbf{ 13,554,762} \\ \hline 
\end{tabular}
}
\end{table}

\smallskip
\noindent\textbf{Static Analysis Tools Warnings Detected in our projects.} We obtained results only for 47 projects out of the 112 contained in the Qualitas Corpus dataset, applying the warnings defined by Better Code Hub, Checkstyle, Coverity Scan, Findbugs, PMD, and SonarQube. Unfortunately, the used versions of  Better Code Hub and Coverity Scan  were not able to analyze all the dataset. So, we considered only the projects analyzed by all the six tools. 

\begin{table}[h]
    \centering
    \footnotesize
    \caption{\textbf{RQ$_1$.} Detection Capability and Detected Warnings in the 47 projects.}
\label{tab:OverallResults}
\resizebox{1\linewidth}{!}{%
    \begin{tabular}{l|l|l|l|l|l|l|l} \hline 
\multirow{2}{*}{\textbf{Tool}} & \multicolumn{3}{c|}{\textbf{Detection Capability}}  & \multicolumn{4}{c}{\textbf{Detected warning}}   \\ \cline{2-8}
& \textbf{\# rule } & \textbf{type} & \textbf{severity} & \textbf{\# rule }& \textbf{\# occurrences }& \textbf{type} & \textbf{severity}\\ \hline 

Better Code Hub & 10 & 3 & 3 & 8 & 27,888 & 3 & 3\\
Checkstyle & 173 & 14 & 4 & 88 & 9,686,813 & 12 & 2\\
Coverity &  &  &  & 130 & 7,431 & 26 & 3\\
Findbugs & 424 & 9 & 4 & 255 & 33,704 & 9 & 3\\
PMD & 305 & 8 & 5 & 275 & 3,380,493 & 7 & 5\\
SonarQube & 413 & 3 & 5 & 180 & 418,433 & 3 & 5\\\hline 

\end{tabular}
}
\end{table}

\begin{table}[h]
\tiny
    \centering
        \caption{\textbf{RQ$_1$.} The top-10 warnings detected by Better Code Hub, Checkstyle, Coverity Scan, FindBugs, PMD, and SonarQube.}
    \label{tab:warningsDetected}
    \begin{tabular}{p{0.7cm}|p{9.5cm}|p{0.8cm}} \hline 
\textbf{Id} & \textbf{Better Code Hub Detected Rule} & \textbf{\#}  \\ \hline 
        &WRITE\_CLEAN\_CODE	&16,055\\
&WRITE\_CODE\_ONCE	&14,692\\
&WRITE\_SHORT\_UNITS	& 6,510\\
&AUTOMATE\_TESTS	& 6,475 \\
&WRITE\_SIMPLE\_UNITS & 6,362\\ 
&SMALL\_UNIT\_INTERFACES	&6,352\\
&SEPARATE\_CONCERNS\_IN\_MODULES	&5,880\\
&COUPLE\_ARCHITECTURE\_COMPONENTS\_LOOSELY & 2,807 \\\hline  \hline 

\textbf{Id} & \textbf{Checkstyle Detected Rule} & \textbf{\#}  \\ \hline
&IndentationCheck &3,997,581\\
&FileTabCharacterCheck &2,406,876\\
&WhitespaceAroundCheck &865,339\\
&LeftCurlyCheck &757,512\\
&LineLengthCheck &703,429\\
&RegexpSinglelineCheck &590,020\\
&FinalParametersCheck &406,331\\
&ParenPadCheck &333,007\\
&NeedBracesCheck &245,110\\
&MagicNumberCheck &223,398\\
\hline \hline 

\textbf{Id} & \textbf{Coverty Scan Detected Rule} & \textbf{\#}  \\ \hline 
&Dereference null return value	&	1,360	\\
&Dm: Dubious method used	&	689	\\
&Unguarded read	&	556	\\
&Explicit null dereferenced	&	514	\\
&Resource leak on an exceptional path	&	494	\\
&Dereference after null check	&	334	\\
&Resource leak	&	301	\\
&DLS: Dead local store	&	293	\\
&Missing call to superclass	&	242	\\
&Se: Incorrect definition of serializable class	&	224	\\ \hline \hline 
\textbf{Id} & \textbf{FindBugs Detected Rule} & \textbf{\#}  \\ \hline 
& BC\_UNCONFIRMED\_CAST & 2,840 \\
& DM\_NUMBER\_CTOR & 2,557 \\
& BC\_UNCONFIRMED\_CAST\_OF\_RETURN\_VALUE & 2,424 \\
& DM\_DEFAULT\_ENCODING & 1,946 \\
& RCN\_REDUNDANT\_NULLCHECK\_OF\_NONNULL\_VALUE & 1,544 \\
& DLS\_DEAD\_LOCAL\_STORE & 1,281 \\
& DM\_FP\_NUMBER\_CTOR & 959 \\
& SE\_NO\_SERIALVERSIONID & 944 \\
& REC\_CATCH\_EXCEPTION & 887 \\
& SE\_BAD\_FIELD & 878 \\
\hline \hline 

\textbf{Id} & \textbf{PMD Detected Rule} & \textbf{\#}  \\ \hline
& LawOfDemeter & 505,947 \\
& MethodArgumentCouldBeFinal & 374,159 \\
& CommentRequired & 368,177 \\
& LocalVariableCouldBeFinal & 341,240 \\
& CommentSize & 153,464 \\
& DataflowAnomalyAnalysis & 152,681 \\
& ShortVariable & 136,162 \\
& UselessParentheses & 128,682 \\
& BeanMembersShouldSerialize & 111,400 \\
& ControlStatementBraces & 110,241 \\
\hline \hline 

\textbf{Id} & \textbf{SonarQube Detected Rule} & \textbf{\#}  \\ \hline 
S1213	&	The members of an interface declaration or class should appear in a pre-defined order & 30,888	\\ 
S125	& Sections of code should not be ''commented out'' &	30,336	\\
S00122	&	Statements should be on separate lines & 26,072	\\
S00116	&	Field names should comply with a naming convention	&25,449	\\
S00117	&	Local variable and method parameter names should comply with a naming convention 	& 23,497	\\
S1166	&	Exception handlers should preserve the original exceptions	&21,150	\\
S106	&	Standard outputs should not be used directly to log anything	&19,713	\\
S1192	&	String literals should not be duplicated	& 19,508	\\
S134	&	Control flow statements ''if'',''for'',''while'',''switch'' and ''try'' should not be nested too deeply 	&17,654	\\
S1132	& Strings literals should be placed on the left side when checking for equality		&13,576	\\\hline
    \end{tabular}
\end{table} 

In total, the projects were infected by \textbf{936} warnings violated \textbf{13,554,762 times}. 
8 (out of 10) warnings were detected by Better Code Hub 27,888 times, 88 (out of 173) warnings were detected by Checkstyle 9,686,813 times, 130 warnings were detected by Coverity Scan 7,431 times, 255 (out of 424) warnings were detected by Findbugs 33,704 time, 275 (out of 305) warnings were detected by PMD 3,380,493 times, and 180 (out of 413) warnings were detected by SonarQube 418.433 times (Table~\ref{tab:Diffusion} and Table~\ref{tab:OverallResults}). It is important to note that in Table~\ref{tab:OverallResults}, the detection capability is empty for Coverity. As mentioned earlier, the full detection capability is only provided to clients and not on the public API. We also computed how often warnings were violated by grouping them based on type and severity. The full results of this additional analysis are reported in our replication package\footref{package}.

Given the scope of warnings that were detected, our projects were affected by all warnings that are detectable by Better Code Hub and by some warnings that are detectable by Coverity Scan and SonarQube (Table~\ref{tab:warningsDetected}). 
For the sake of readability, we report only the Top-10 warnings detected in our projects by the six tools. The complete list is available in the replication package\footref{package}.

\smallskip
\roundedbox{Our analysis provided a mapping of the warnings that currently available tools can identify in software projects. Overall, the amount of warnings detected by the six automated static analysis tools is significant (\textbf{936} warnings detected \textbf{13,554,762 times}); hence, we can proceed with the analysis of the remaining RQ$_s$.} 

\begin{table}[h]
\footnotesize
\caption{\textbf{RQ$_2$.} Rule pairs that overlap at the ``class-level''.}
\label{tab:warningspairsClass}
    \centering
    \begin{tabular}{p{3.2cm}|p{2.2cm}|p{3cm}|p{0.8cm}} \hline
  \textbf{Rule pairs} & \textbf{\# occurrences} & \textbf{\# max occurrences} & \textbf{\%} \\ \hline 
Checkstyle - PMD & 4,872 & 3,380,493 & 0.144\\ 
SonarQube - PMD & 4,126 & 418,433 & 0.98 \\
Findbugs - PMD & 3,161 & 33,704 & 9.378 \\
SonarQube - Checkstyle & 1,495 & 418,433 & 0.357\\
Findbugs - Checkstyle & 1,265 & 33,704 & 3.753\\
BCH-PMD & 1,017 & 27,888 & 3.646\\
SonarQube - Findbugs & 849 & 33,704 & 2.518\\
BCH-SonarQube    &  517 & 27,888 & 1.853\\
BCH-Checkstyle & 440 & 27,888 & 1.577\\
BCH-Findbugs    &  235 & 27,888 & 0.842\\ 
Coverity - BCH & 117 & 7,431 & 1.574\\ 
Coverity - Checkstyle & 128 & 7,431 & 1.723\\ 
Coverity - Findbugs & 120 & 7,431 & 1.615\\ 
Coverity - PMD & 128 & 7,431 & 1.723\\ 
Coverity - SonarQube & 128 & 7,431 & 1.723\\  \hline 
\textbf{Total}  & \textbf{18,598} & \textbf{4,457,178} & \textbf{0.417} \\ \hline 
\end{tabular}
\end{table}

\noindent\textbf{RQ$_2$. What is the agreement among different Static Analysis Tools?} Our second research question focused on the analysis of the agreement between the static analysis tools. 

\smallskip
\textbf{Agreement based on the overlapping at ``class-level''.} In order to include Coverity Scan in this analysis, we first evaluated the detection agreement at ``class-level'', considering each class where the warnings detected by the other five tools overlapped at 100\% and where at least one warning of Coverity Scan was violated in the same class. 

To calculate the percentage of warnings pairs (columns ``\%'', Table~\ref{tab:warningspairsClass}) that appear together, we checked the occurrences of both tools in our projects, then we considered only the minimum value. For example, in Table~\ref{tab:warningspairsClass}, calculating the percentage between Checkstyle - PMD warning pairs, we have 9,686,813 warnings Checkstyle detected and 3,380,493 PMD ones detected. The combination of these warnings should be maximum 3,380,493 (the minimum value between the two). We calculated the percentage considering the column ``\# occorrences'' and the column ``\# possible occorrences''.  

\begin{table}[H]
\footnotesize
\caption{\textbf{RQ$_2$.} Rule pairs that overlap at the 100\% threshold considering the  ``line-level''.}
\label{tab:warningspairs100}
    \centering
    \begin{tabular}{p{3.2cm}|p{2.2cm}|p{3.5cm}|p{0.8cm}} \hline
   \textbf{Rule pairs} & \textbf{\# occurrences} & \textbf{\# possible occurrences} & \textbf{\%} \\ \hline 
Checkstyle - PMD & 4,872 & 3,380,493 & 0.144\\ 
SonarQube - PMD & 4,126 & 418,433 & 0.98 \\
Findbugs - PMD & 3,161 & 33,704 & 9.378 \\
SonarQube - Checkstyle & 1,495 & 418,433 & 0.357\\
Findbugs - Checkstyle & 1,265 & 33,704 & 3.753\\
BCH-PMD & 1,017 & 27,888 & 3.646\\
SonarQube - Findbugs & 849 & 33,704 & 2.518\\
BCH-SonarQube    &  517 & 27,888 & 1.853\\
BCH-CheckSyle & 440 & 27,888 & 1.577\\
BCH-Findbugs    &  235 & 27,888 & 0.842\\ \hline  
\textbf{Total} & \textbf{17,977} & \textbf{4,430,023} & \textbf{0.4\%} \\ \hline 
\end{tabular}
\end{table}

The warnings overlap at the ``class-level'' is always low, as reported in Table~\ref{tab:warningspairsClass}. This means that a piece of code violated by a warning detected by one tool is almost never violated by another warning detected by another tool. In the best case (Table~\ref{tab:warningspairsClass}), only 9.378\% of the possible warning (Findbugs-PMD). 
Moreover, we did find no warnings pair at ``class-level'' considering more that two tools (e.g. Checkstyle-Findbugs-PMD). 

For each warnings pair we computed the detection agreement at class level. For the sake of readability, we report these results in Appendix A. Specifically, the three tables (Table~\ref{tab:AllToghether1}, Table~\ref{tab:AllToghether2}, and Table~\ref{tab:AllToghether3}) overview the detection agreement of each warning pair, according to the procedure described in Section~\ref{DataAnalysis}. As further explained in the appendix, for reasoning of space we only showed the 10 most recurrent pairs, putting the full set of results in our replication package~\footref{package}.
In these tables, the third and fourth columns (eg. ``\#BCH pairs'' and ``\# CHS pairs'', Table~\ref{tab:AllToghether1}) report how many times a warning instance from a tool exists with another one. The tools have separate values for the number of co-occurrences as the number of instances differs as well, for example, could be that a large rule contains several instances of the comparison rule. Then for other rule this counts as one co-occurrence while for the other rule each included rule grows the number. This makes sure the agreement is between 0 and 1 for both rules. The remaining two columns report the agreement of each tools considered in the warning pairs (eg. ``\#BCH Agr.'' and ``\# CHS Agr.'', Table~\ref{tab:AllToghether1}). 
Results showed for all the warning pairs that the agreement at ``class-level'' is very low, as none of the most recurrent warning pairs agree well. The results also highlighted the difference in the granularity of the warnings.

\smallskip
\textbf{Agreement based on the overlapping at the ``line-level".} Since we cannot compare at ``line-level'' the warnings detected by Coverity Scan, we could only consider the remaining five static analysis tools. Using the bottom-up approach (Figure~\ref{fig:thresholds}), several rule pairs were found according to the 100\%, 90\%, 80\%, and 70\% thresholds. Using the threshold of \textbf{100\%} which indicates that a rule completely resides within the reference rule,  we found \textbf{17,977} \textbf{rule pairs}, as reported in Table~\ref{tab:warningspairs100}. Using the thresholds of 90\%, 80\%, and 70\% the following rule pairs were found respectively: 17,985, 18,004, and 18,025 (Table~\ref{tab:warningspairs90}). These warnings resided partially within the reference rule. 

\begin{table}[H]
\footnotesize
\caption{\textbf{RQ$_2$.} Rule pairs that overlap at the different thresholds, i.e., 90/80/70\%, considering the ``line-level''.} 
\label{tab:warningspairs90}
    \centering
    \begin{tabular}{p{3.1cm}|p{1.8cm}|p{1.8cm}|p{1.8cm}|p{1.4cm}} \hline
\multirow{3}{*}{\textbf{Rule pairs}} & \multicolumn{4}{c}{\textbf{\# occurrences}}   \\ \cline{2-5}
 & \textbf{90\%} & \textbf{80\%} & \textbf{70\%} &  \multirow{2}{*}{\textbf{possible}}  \\\cline{2-4}
 & \textbf{\#(\%)} & \textbf{\#(\%)}& \textbf{\#(\%)} & \\ \hline 
Checkstyle - PMD & 4,872 (0.144) & 4,874 (0.144)& 4,876 (0.144)& 3,380,493 \\ 
SonarQube - PMD & 4,126 (0.986)& 4,130 (0.987)& 4,139 (0.989)& 418,433  \\
Findbugs - PMD & 3,167 (9.39)& 3,173 (9.41)& 3,173 (9.41)& 33,704  \\
SonarQube - Checkstyle & 1,495 (0.357)& 1,496 (0.357)& 1,496 (0.357)& 418,433  \\
Findbugs - Checkstyle & 1,265 (3.753)& 1,265 (3.753)& 1,265 (3.753) & 33,704  \\
BCH-PMD & 1,017 (3.646)& 1,019 (3.646)& 1,024 (3.647)& 27,888  \\
SonarQube - Findbugs & 849 (2.519) & 849 (2.519)& 849 (2.519)& 33,704  \\
BCH-SonarQube    &  517 (1.853) & 521 (1.868) & 522 (1.868) & 27,888  \\
BCH-CheckSyle & 440 (1.577)& 440 (1.577)& 441 (1.578)& 27,888 \\
BCH-Findbugs    &  237 (0.849) & 237 (0.849)& 240 (0.860)& 27,888 \\  \hline 
\textbf{Total} & \textbf{18,004} \textbf{(0.4)} & \textbf{18,004}  \textbf{(0.4)} & \textbf{18,025}  \textbf{(0.4)} & \textbf{4,430,023}  \\ \hline 
\end{tabular}
\end{table}

Similarly to what happened with the agreement at ``class-level'', it is important to note that the overlap at the ``line-level'' is always low. Results show that, also in this case, only 9.378\% of the possible rule occurrences are detected in the same line by the same two tools (Findbugs and PMD). 
In addition, also in this case we did find no pair warnings at ``line-level'' considering more that two tools (e.g. Checkstyle-Findbugs-PMD).

When considering the agreement for each warning pair at ``line-level`, we could not obtain any result because of computational reasons. Indeed, the analysis at line-level of 936 warning types that have been violated 13,554,762 times would have required a prohibitively expensive amount of time/space---according to our estimations, it would have been taken up to 1.5 years---and, therefore, we preferred excluding it. 

\smallskip
\roundedbox{The warnings overlapping among the different tools is very low (less than 0.4\%). The warning pairs Checkstyle - PMD as the lowest overlap (0.144\%) and  Findbugs - PMD the highest one (9.378\%). Consequently also the detection agreement is very low.} 

\smallskip
\noindent\textbf{RQ$_3$. What is the precision of the static analysis tools?}
\label{results:rq3}

In the context of our last research question, we focused on the precision of the static analysis tools when employed for TD detection. Table \ref{tab:precision} reports the results of our manual analyses. As shown, the precision of most tools is quite low, e.g., SonarQube has a precision of 18\%, with the only exception of CheckStyle whose precision is equal to 86\%.

\begin{table}[H]
\footnotesize
    \centering
        \caption{\textbf{RQ$_3$.} Precision of the considered SATs over the manually validated sample set of warnings.}
    \label{tab:precision}
    \begin{tabular}{l|r|r|r} \hline 
\textbf{SAT} & \textbf{\# warnings} & \textbf{\# True Positives} & \textbf{Precision} \\ \hline
    Better Code Hub & 375 & 109 & 29\% \\
    Checkstyle & 384 & 330 & 86\% \\
    Coverity Scan & 367 & 136 & 37\% \\
    Findbugs & 379 & 217 & 57\% \\
    PMD & 384 & 199 & 52\% \\
    SonarQube & 384 & 69 & 18\% \\
 \hline 
    \end{tabular}
\end{table}

In general, based on our findings, we can first corroborate previous findings in the field \cite{antunes2009comparing,johnson2013don,mclean2012comparing} and the observations reported by Johnson et al. \cite{johnson2013don}, who found through semi-structured interviews with developers that the presence of false positives represents one of the main issues that developers face when using static analysis tools in practice. With respect to the qualitative insights obtained by interviewing developers \cite{johnson2013don}, our work concretely quantifies the capabilities of the considered static analysis tools. 

Looking deeper into the results, we could delineate some interesting discussion points. First, we found that for Better Code Hub and Coverity Scan almost two thirds of the recommendations represented false alarms, while the lowest performance was achieved by SonarQube. The poor precision of the tools is likely due to the high sensitivity of the warnings adopted to search for potential issues in the source code, e.g., threshold values that are too low lead to the identification of false positive TD items. This is especially true in the case of SonarQube: In our dataset, it outputs an average of 47.4 violations per source code class, often detecting potential TD in the code too hastily. 

A slightly different discussion is the one related to the other three static analysis tools, namely PMD, Findbugs, and Checkstyle. 

As for the former, we noticed that it typically fails when raising warnings related to naming conventions. For instance, this is the case of the \textsl{'AbstractName'} warning: it suggests the developer that an abstract class should contain the term \texttt{Abstract} in the name. In our validation, we discovered that in several cases the recommendation was wrong because the contribution guidelines established by developers explicitly indicated alternative naming conventions. A similar problem was found when considering FindBugs. The precision of the tool is 57\% and, hence, almost half of the warnings were labeled as false positives. In this case, one of the most problematic cases was related to the \textsl{'BC\_UNCONFIRMED\_CAST'} warnings: these are raised when a cast is unchecked and not all instances of the type casted from can be cast to the type it is being cast to. In most cases, these warnings have been labeled as false positives because, despite casts were formally unchecked, they were still correct by design, i.e., the casts could not fail anyway because developers have implicitly ensured that all of them were correct. 

Finally, Checkstyle was the static analysis tool having the highest precision, i.e., 86\%. When validating the instances output by the tool, we realized that the warnings raised are related to pretty simple checks in source code that cannot be considered false positives, yet do not influence too much the functioning of the source code. To make the reasoning clearer, let consider the case of the \textsl{'IndentationCheck'} warning: as the name suggests, it is raised when the indentation of the code does not respect the standards of the project. In our sample, these warnings were all true positives, hence contributing to the increase of the precision value. However, the implementation of these recommendations would improve the documentation of the source code but not dealing with possible defects or vulnerabilities. As such, we claim that the adoption of Checkstyle would be ideal when used in combination with additional static analysis tools.

To broaden the scope of the discussion, the poor performance achieved by the considered tools reinforces the preliminary research efforts to devise approaches for the automatic/adaptive configuration of static analysis tools~\cite{nadi2014mining,di2017dynamic} as well as for the automatic derivation of proper thresholds to use when locating the presence of design issues in source code~\cite{aniche2016satt,fontana2015automatic}. It might indeed be possible that the integration of those approaches into the inner workings of the currently available static analysis tools could lead to a reduction of the number of false positive items. In addition, our findings also suggest that the current static analysis tools should not limit themselves to the analysis of source code but, for instance, complementing it with additional resources like naming conventions actually in place in the target software system.

\smallskip
\roundedbox{
Most of the considered SATs suffer from a high number of false positive warnings, and their precision ranges between 18\% and 57\%. The only expection is Checkstyle (precision=86\%), even though most of the warnings it raises are related to documentation issues rather than functional problems and, as such, its adoption should be complemented with other static analysis tools.
}

\section{Discussion and Implications}
\label{Discussion}

The results of our study provide a number of insights that can be used by researchers and tool vendors to improve SATs. Specifically, these are:

\smallskip
\textbf{There is no silver bullet.} According to the results obtained in our study, and specifically for \textbf{RQ$_1$}, different SAT warnings are able to cover different issues, and can therefore find different forms of source code quality problems: Hence, we can claim that \emph{there is no silver bullet that is able to guarantee source code quality assessment on its own}. On the one hand, this finding highlights that practitioners interested in detecting quality issues in their source code might want to combine multiple SATs to find a larger variety of problems. On the other hand, and perhaps more importantly, our results suggest that the research community should have an interest in and be willing to devise more advanced algorithms and techniques, e.g., ensemble methods or meta-models \cite{catolino2019extensive, catolino2019improving,catolino2018enhancing,palomba2017toward}, that can (1) combine the results from different static analysis tools and (2) account for possible overlaps among the rules of different SATs. This would allow the presentation of more complete reports about the code quality status of the software systems to their developers.

\smallskip
\textbf{Learning to deal with false positives.} One of the main findings of our study concerns with the low performance achieved by all static analysis tools in terms of precision of the recommendations provided to developers (\textbf{RQ$_3$}). Our findings represent the first attempt to concretely quantify the capabilities of the considered SATs in the field. Moreover, our study provides two practical implications: (1) It corroborates and triangulates the qualitative observations provided by Johnson et al.~\cite{johnson2013don}, hence confirming that the real usefulness of static analysis tools is threatened by the presence of false positives; (2) it supports the need for more research on how to deal with false positives, and particularly on how to filter likely false alarms~\cite{fontana2015filtering} and how to select/prioritize the warnings to be presented to developers~\cite{kim2007warnings,liang2010automatic,LenarduzziSANER2019}. While some preliminary research efforts on the matter have been made, we believe that more research should be devoted to these aspects.
Finally, our findings may potentially suggest the need for further investigation into the effects of false positives in practice: For example, it may be worthwhile for researchers to study what the maximum number of false positive instances is that developers can deal with, e.g., they should devise a critical mass theory for false positive ASAT warnings~\cite{oliver2001whatever} in order to augment the design of existing tools and the way they present warnings to developers.

\smallskip
\textbf{Complementing static analysis tools.} The findings from our \textbf{RQ$_1$} and \textbf{RQ$_2$} highlight that most of the issues reported by the state-of-the-art static analysis tools are related to rather simple problems, like the writing of shorter units or the automation of software tests. These specific problems could possibly be avoided if current static analysis tools would be complemented with effective tools targeting (1) automated refactoring and (2) automatic test case generation. In other words, our findings support and strongly reinforce the need for a joint research effort among the communities of source code quality improvement and testing, which are called to study possible synergies between them as well as to devise novel approaches and tools that could help practitioners complement the outcome provided by static analysis tools with that of other refactoring and testing tools. For instance, with effective refactoring tools, the number of violations output by SATs would be notably reduced, possibly enabling practitioners to focus on the most serious issues. 

\section{Threats to Validity}
\label{Threats}
A number of factors might have influenced the results reported in our study. This section discusses the main threats to validity and how we mitigated them.

\smallskip
\textbf{Construct Validity}.
Threats in this category concern the relationship between theory and observation. A first aspect is related to the dataset used. In our work, we selected 112 projects from the Qualitas Corpus \cite{Terra2013}, which is one of the most reliable data sources in software engineering research \cite{Tempero2010}. 
Another possible threat relates to the configuration of the SATs employed. None of the considered projects had all the static analysis tools configured and so we had to manually introduce them; in doing so, we relied on the default configuration of the tools since we could not rely on different configurations given directly by the developers of the projects. Nevertheless, it is important to point out that this choice did not influence our analyses: indeed, we were interested in comparing the capabilities of existing tools independently from their practical usage in the considered systems. The problem of configuring the tools therefore does not change the answers to our research questions.

\smallskip
\textbf{Internal Validity}. 
As for potential confounding factors that may have influenced our findings, it is worth mentioning that some issues detected by SonarQube were duplicated: in particular, in some cases the tool reported the same issue violated in the same class multiple times. To mitigate this issue, we manually excluded those cases to avoid interpretation bias; we also went over the rules output by the other static analysis tools employed to check for the presence of duplicates, but we did not find any.

\smallskip
\textbf{External Validity}.
Threats in this category are concerned with the generalization of the results. 
While we cannot claim that our results fully represent every Java project, we considered a large set of projects with different characteristics, domains, sizes, and architectures. This makes us confident of the validity of our results in the field, yet replications conducted in other contexts would be desirable to corroborate the reported findings. 

Another discussion point is related to our decision to focus only on open-source projects. In our case, this was a requirement: we needed to access the code base of the projects in order to configure the static analysis tools. Nevertheless, open-source projects are comparable---in terms of source code quality---to closed-source or industrial applications~\cite{LenarduzziOSS2019}; hence, we are confident that we might have obtained similar results by analyzing different projects. Nevertheless, additional replications would provide further complementary insights and are, therefore, still desirable.

Finally, we limited ourselves to the analysis of Java projects, hence we cannot generalize our results to projects in different programming languages. Therefore, further replications would be useful to corroborate our results.

\smallskip
\textbf{Conclusion Validity}.
With respect to the correctness of the conclusions reached in this study, this has mainly to do with the data analysis processes used. In the context of \textbf{RQ$_1$} and \textbf{RQ$_3$}, we conducted iterative manual analyses in order to build the taxonomy and study the precision of the tools, respectively. While we cannot exclude possible imprecision, we mitigated this threat by involving more than one inspector in each phase, who first conducted independent evaluations that were later merged and discussed. Perhaps more importantly, we made all data used in the study publicly available with the aim of encouraging replicability, other than a further assessment of our results.

In \textbf{RQ$_2$} we proceeded with an automatic mechanism to study the agreement among the tools. As explained in Section \ref{StudySetup}, different static analysis tools might possibly output the same warnings in slightly different positions of the source code, e.g., highlighting the violation of a rule at two subsequent lines of code. To account for this aspect, we defined thresholds with which we could manage those cases where the same warnings were presented in different locations. In this case, too, we cannot exclude possible imprecision; however, we extensively tested our automated data analysis script. More specifically, we manually validated a subset of rules for which the script indicated an overlap between two tools with the aim of assessing whether it was correct or not. This manual validation was conducted by one of the authors of this paper, who took into account a random sample of 300 candidate overlapping rules. In this sample, the author could not find any false positives, meaning that our script correctly identified the agreement among tools. This further analysis makes us confident of the validity of the findings reported for \textbf{RQ$_2$}.

\section{Related Work}
\label{RW}
Automated Static Analysis Tools (ASATs) are getting more popular~\cite{vassallo2019developers, LenarduzziSEDA2019} as they are becoming easier to use~\cite{zampetti2017open}. The use of static analysis tools has been studied by several researchers in the last years~\cite{Wagner2005, Nagappan2005, Zheng2006, Nanda2010}. 
In this section, we report the relevant work on static analysis tools focusing on their usage~\cite{SaarimakiTechDebt2019, LenarduzziMALTESQUE2019, LenarduzziJSS2019}, warnings and the detected problems~\cite{Flanagan2002, Heckman2011, Beller2016}. 

Developers can use ASATs, such as SonarQube\footref{Sonar} and CheckStyle\footnote{\label{Checkstyle} https://checkstyle.sourceforge.io/}, to evaluate software source code, finding anomalies of various kinds in the code~\cite{Rutar2004, Tomas2013}.
Moreover, ASATs are widely adopted in many research studies in order to evaluate the code quality~\cite{johnson2013don, Schnappinger2019, Marcilio2019} and identify issues in the code~\cite{SaarimakiTechDebt2019, LenarduzziMALTESQUE2019, LenarduzziJSS2019}. Some studies demonstrated that some rules detected by ASATs can be effective for identifying issues in the code~\cite{Zheng2006, LenarduzziSANER2019, LenarduzziJSS2019}. However, evaluating the performance in defect prediction, results are discordant comparing different tools (e.g. FindBugs\footref{Findbugs} and PMD\footref{PMD})~\cite{Rahman2014}. 

Rutar et al.~\cite{Rutar2004} compared five bug-finding tools for Java (Bandera\footnote{http://bandera.projects.cs.ksu.edu/}, ESC/Java2\footnote{https://kindsoftware.com/products/opensource/ESCJava2/}, FindBugs\footnote{\label{Findbugs} http://findbugs.sourceforge.net/}, JLint\footnote{\label{JLint} http://jlint.sourceforge.net/}, and PMD\footnote{\label{PMD} https://pmd.github.io/}), that use syntactic bug pattern detection, on five projects, including JBoss 3.2.3\footnote{http://www.jboss.org/} and Apache Tomcat 5.019\footnote{http://jakarta.apache.org/tomcat}. They focused on the different warnings (also called rules) provided by each tool, and their results demonstrate some overlaps among the types of errors detected, which may be due to the fact that each tool applies different trade-offs to generate false positives and false negatives. Overall, they stated that warnings provided by the different tools are not correlated with each other. Complementing the work by Rutar et al.~\cite{Rutar2004}, we calculated the agreement of ASATs on TD identification. In addition, we investigated the precision with which these tools output warnings. Finally, we also investigated the types of TD items that can actually be detected by existing ASATs.

Tomas et al.~\cite{Tomas2013} performed a comparative analysis by means of a systematic literature review. In total, they compared 16 Java code static analysis tools, including JDepend\footnote{https://github.com/clarkware/jdepend},  Findbugs\footref{Findbugs}, PMD\footref{PMD}, and  SonarQube\footref{Sonar}. They focused on internal quality metrics of a software product and software tools of static code analysis that automate measurement of these metrics. As results, they reported the tools' detection strategies and what they detect. For instance, most of them automate the calculation of internal quality metrics, the most common ones being code smells, complexity, and code size~\cite{Tomas2013}. However, they did not investigate agreement between the tools' detection rules. 

Avgeriou et al.~\cite{Avgeriou2020} identified the available static analysis tools for the Technical Debt detection. They compared features and popularity of nine tools investigating also the empirical evidence on their validity. 
Results can help practitioners and developers to select the suitable tool against the other ones according to the measured information that satisfied better their needs. However, they did not evaluate their agreement and precision in the detection. 

Focusing on developers' perception on the usage of static analysis tools, ASATs can help to find bugs~\cite{johnson2013don}. However, developers are not sure about the usefulness of the rules ~\cite{TaibiIST2017, Vassallo2018, Sadowski2018}, they do pay attention to different rules categories and priorities and remove violations related to rules with high severity~\cite{Vassallo2018} in order to avoid the possible risk of faults~\cite{TaibiIST2017}.
Moreover, false positives and the way in which the warnings are presented, among other things, are barriers to their wider adoption~\cite{johnson2013don}. Some studies highlighted the need to reduce the number of detectable rules~\cite{Muske2018, Bodden2018} or summarize them based on similarities~\cite{Vassallo2018}. 
ASATs are able to detect many defects in the code. However, some tools  do not capture all the possible defect even if they could be detected by the tools~\cite{Thung2015}. Even if some studies since the beginning of 2010 highlighted the need to better clarify the precision of the tools, differentiating false positives from actionable rules~\cite{liang2010automatic, Ruthruff2008}, many studies deal with the many false positives produced by different tools, such as FindBugs\footref{Findbugs}~\cite{Thung2015, Ayewah2010, Ayewah2008}, JLint\footref{JLint}, PMD\footref{PMD}, CheckStyle\footref{Checkstyle}, and JCSC\footnote{http://jcsc.sourceforge.net}~\cite{Thung2015}.

At the best of our knowledge, our work is the first that investigate in details which source quality problems can actually be detected by the available tools, trying to make a comparison based on the description, what is their agreement, and  what is the precision of their recommendations. 

\section{Conclusion}
\label{Conclusion}
In this paper, we performed a large-scale comparison of six popular Static Analysis Tools (Better Code Hub, CheckStyle, Coverity Scan, Findgugs, PMD, and SonarQube) with respect to the detection of static analysis warnings. We analyzed 47 Java projects from the Qualitas Corpus dataset, and derived similar warnings that can be detected by the tools. We also compared their detection agreement at ``line-level'' and ``class-level'', and manually analyzed their precision. 
To sum up, the contributions of this paper are:

\begin{enumerate}
    \item A comparison of the warnings that can be detected by the tools (taxonomy), which may be useful for researchers and tool vendors to understand which warnings should be considered during refactoring; 
    
    \item An analysis of the agreement among the tools, which can inform tool vendors about the limitations of the current solutions available the market;
    
    \item The first quantification of the precision of six static analysis tools (Better Code Hub, CheckStyle, Coverity Scan, Findgugs, PMD, and SonarQube). 
    
\end{enumerate}

Our future work includes an extension of this study with the evaluation of the recall, and the in-vivo assessment of the tools.

\bibliographystyle{spbasic}   
\bibliography{sample}

\newpage
\appendix 
\section*{Appendix A}
\label{Appendix}
In the following, we report more results achieved in the context of \textbf{RQ$_2$}. For each warnings pair we computed the detection agreement at class level as reported in Table~\ref{tab:AllToghether1}, Table~\ref{tab:AllToghether2}, and Table~\ref{tab:AllToghether3}, according to the process described in Section~\ref{DataAnalysis}. It is worth remarking that, for the sake of readability, we only show the 10 most recurrent pairs. The results for the remaining thresholds are reported in the replication package\footref{package}.

\begin{table}[H]
 \caption{\textbf{RQ$_2$.} The 10 most recurrent rule pairs detected in the same class by the considered SATs and their corresponding agreement values.}
 \label{tab:AllToghether3}
 \centering
 \tiny
 \begin{tabular}{p{2.7cm}| p{5cm}|p{0.45cm}|p{0.45cm}|p{0.45cm}|p{0.45cm}}

\toprule
\textbf{SonarQube} &  \textbf{FindBugs} &  \textbf{\# SQ pairs }&  \textbf{\# FB pairs} &  \textbf{SQ Agr.} &  \textbf{FB Agr.} \\
\midrule
CommentedOutCodeLine &  RV\_RETURN\_VALUE\_OF\_PUTIFABSENT\_IGNORED & 2 & 1 & 0.000 & 1.000 \\
S1155 &  RV\_RETURN\_VALUE\_OF\_PUTIFABSENT\_IGNORED & 1 & 1 & 0.000 & 1.000 \\
S1186 &  RV\_RETURN\_VALUE\_OF\_PUTIFABSENT\_IGNORED & 1 & 1 & 0.000 & 1.000 \\
S135 &  RV\_RETURN\_VALUE\_OF\_PUTIFABSENT\_IGNORED & 3 & 1 & 0.001 & 1.000 \\
S1312 &  RV\_RETURN\_VALUE\_OF\_PUTIFABSENT\_IGNORED & 1 & 1 & 0.000 & 1.000 \\
complex\_class &  RV\_RETURN\_VALUE\_OF\_PUTIFABSENT\_IGNORED & 1 & 1 & 0.000 & 1.000 \\
S1195 &DM\_DEFAULT\_ENCODING & 1 & 1 & 1.000 & 0.001 \\
S1195 &EI\_EXPOSE\_REP & 1 & 1 & 1.000 & 0.001 \\
S1301 & LG\_LOST\_LOGGER\_DUE\_TO\_WEAK\_REFERENCE & 1 & 1 & 0.001 & 1.000 \\
S1148 &FI\_NULLIFY\_SUPER & 1 & 1 & 0.000 & 1.000 \\
\bottomrule

\toprule
\textbf{SonarQube} &  \textbf{PMD} &  \textbf{\# SQ pairs} &  \textbf{\# PMD pairs} &  \textbf{SQ Agr.} &  \textbf{PMD Agr.} \\
\midrule
 S1192 &  FinalizeOnlyCallsSuperFinalize & 1 &  1 & 0.000 &  1.000 \\
 S2110 &  JUnit4SuitesShouldUseSuiteAnnotation & 1 &  1 & 1.000 &  0.001 \\
 S2110 &AvoidCatchingGenericException & 1 &  5 & 1.000 &  0.000 \\
 S2110 & AvoidCatchingNPE & 1 &  4 & 1.000 &  0.011 \\
 S2110 &AvoidPrintStackTrace & 1 &  1 & 1.000 &  0.000 \\
 S2110 & CloseResource & 1 &  2 & 1.000 &  0.000 \\
 S2110 &CommentSize & 1 &  2 & 1.000 &  0.000 \\
 S2110 &DataflowAnomalyAnalysis & 1 &  5 & 1.000 &  0.000 \\
 S2110 &JUnit4TestShouldUseBeforeAnnotation & 1 &  1 & 1.000 &  0.001 \\
 S2110 & ShortVariable & 1 & 25 & 1.000 &  0.000 \\
\bottomrule

\end{tabular}
\begin{tablenotes}
    \scriptsize
    \centering
    \item \textit{BCH} means Better Code Hub, and \textit{SQ} means SonarQube, while \textit{Ag.} means Agreement
\end{tablenotes}
\end{table}

\newpage
\begin{table}[H]
 \centering
 \tiny
  \caption{\textbf{RQ$_2$.} The 10 most recurrent rule pairs detected in the same class by the considered SATs and their corresponding agreement values.}
 \label{tab:AllToghether1}
 \begin{tabular}{p{4cm}| p{4.5cm}|p{0.45cm}|p{0.45cm}|p{0.45cm}|p{0.45cm}}
 \toprule
 \textbf{BetterCodeHub} & \textbf{ CheckStyle} &  \textbf{\# BCH pairs} &  \textbf{\# CHS pairs} &  \textbf{BCH Agr.} & \textbf{CHS Agr.} \\
 \midrule
  WRITE\_SIMPLE\_UNITS &  AvoidEscapedUnicodeCharactersCheck &  1 & 29859 &  0.000 &  0.529 \\
WRITE\_SHORT\_UNITS &  AvoidEscapedUnicodeCharactersCheck &  1 & 29859 &  0.000 &  0.529 \\
  WRITE\_CODE\_ONCE &  AvoidEscapedUnicodeCharactersCheck &  1 & 29859 &  0.000 &  0.529 \\
  WRITE\_SIMPLE\_UNITS & OperatorWrapCheck &  1 & 60694 &  0.000 &  0.312 \\
WRITE\_SHORT\_UNITS & OperatorWrapCheck &  1 & 60694 &  0.000 &  0.312 \\
  WRITE\_CODE\_ONCE & OperatorWrapCheck &  1 & 60694 &  0.000 &  0.312 \\
WRITE\_SHORT\_UNITS &IllegalTokenTextCheck &  4 &418 &  0.001 &  0.306 \\
  WRITE\_SIMPLE\_UNITS &IllegalTokenTextCheck &  4 &418 &  0.001 &  0.306 \\
AUTOMATE\_TESTS &IllegalTokenTextCheck &  2 &418 &  0.000 &  0.306 \\
  WRITE\_CODE\_ONCE &AtclauseOrderCheck & 30 &210 &  0.002 &  0.306 \\
 \bottomrule
 
 \toprule
 \textbf{BetterCodeHub} &  \textbf{CoverityScan} &  \textbf{\# BCH pairs} &  \textbf{\# CS pairs} &  \textbf{BCH Agr.} &  \textbf{CS Agr.} \\
 \midrule
 AUTOMATE\_TESTS &Exception leaked to user interface &  2 & 1 &  0.000 & 1.000 \\
 SEPARATE\_CONCERNS\_IN\_MODULES &  Unsafe reflection &  2 & 2 &  0.000 & 1.000 \\
 COUPLE\_ARCHITECTURE\_COMPONENTS\_LOOSELY &  AT: Possible atomicity violation &  2 & 1 &  0.001 & 1.000 \\
  WRITE\_CLEAN\_CODE &  Use of hard-coded cryptographic key & 20 & 1 &  0.001 & 1.000 \\
 WRITE\_SIMPLE\_UNITS &Exception leaked to user interface &  2 & 1 &  0.000 & 1.000 \\
 WRITE\_SHORT\_UNITS &Exception leaked to user interface &  2 & 1 &  0.000 & 1.000 \\
WRITE\_CODE\_ONCE &Exception leaked to user interface &  2 & 1 &  0.000 & 1.000 \\
 WRITE\_SHORT\_UNITS &  Unsafe reflection &  2 & 2 &  0.000 & 1.000 \\
 WRITE\_SIMPLE\_UNITS &  Unsafe reflection &  2 & 2 &  0.000 & 1.000 \\
 AUTOMATE\_TESTS &Dead default in switch &  2 & 1 &  0.000 & 1.000 \\
 \bottomrule

\toprule
\textbf{BetterCodeHub} & \textbf{FindBugs} &  \textbf{\# BCH pairs} & \textbf{\# FB pairs} &  \textbf{BCH Agr.} & \textbf{ FB Agr.} \\
\midrule
WRITE\_CODE\_ONCE &ICAST\_BAD\_SHIFT\_AMOUNT & 12 & 8 &  0.001 & 1.000 \\
WRITE\_SIMPLE\_UNITS &  SA\_LOCAL\_SELF\_ASSIGNMENT\_INSTEAD\_OF\_FIELD &  2 & 1 &  0.000 & 1.000 \\
AUTOMATE\_TESTS & FI\_MISSING\_SUPER\_CALL &  3 & 1 &  0.000 & 1.000 \\
SMALL\_UNIT\_INTERFACES &ICAST\_BAD\_SHIFT\_AMOUNT & 12 & 8 &  0.002 & 1.000 \\
WRITE\_SHORT\_UNITS &INT\_VACUOUS\_BIT\_OPERATION &  6 & 1 &  0.001 & 1.000 \\
WRITE\_SHORT\_UNITS &  RV\_RETURN\_VALUE\_OF\_PUTIFABSENT\_IGNORED &  4 & 1 &  0.001 & 1.000 \\
SEPARATE\_CONCERNS\_IN\_MODULES & EQ\_COMPARING\_CLASS\_NAMES &  2 & 1 &  0.000 & 1.000 \\
AUTOMATE\_TESTS & NP\_ALWAYS\_NULL\_EXCEPTION &  2 & 1 &  0.000 & 1.000 \\
WRITE\_CLEAN\_CODE &  RV\_RETURN\_VALUE\_OF\_PUTIFABSENT\_IGNORED &  4 & 1 &  0.000 & 1.000 \\
AUTOMATE\_TESTS &ICAST\_BAD\_SHIFT\_AMOUNT &  3 & 8 &  0.000 & 1.000 \\
\bottomrule

\toprule
\textbf{BetterCodeHub} &  \textbf{PMD} & \textbf{\# BCH pairs} &  \textbf{\# PMD pairs} &  \textbf{BCH Agr.} & \textbf{PMD Agr.} \\
\midrule
SEPARATE\_CONCERNS\_IN\_MODULES &  FinalizeOnlyCallsSuperFinalize &  2 &  1 &  0.000 &  1.000 \\
WRITE\_CODE\_ONCE &  FinalizeOnlyCallsSuperFinalize &  4 &  1 &  0.000 &  1.000 \\
SEPARATE\_CONCERNS\_IN\_MODULES &  AvoidMultipleUnaryOperators &  3 &  2 &  0.001 &  1.000 \\
COUPLE\_ARCHITECTURE\_COMPONENTS\_LOOSELY &  AvoidMultipleUnaryOperators &  3 &  2 &  0.001 &  1.000 \\
SMALL\_UNIT\_INTERFACES &  FinalizeOnlyCallsSuperFinalize &  4 &  1 &  0.001 &  1.000 \\
WRITE\_CLEAN\_CODE &InvalidLogMessageFormat &  2 &  1 &  0.000 &  1.000 \\
AUTOMATE\_TESTS &  FinalizeOnlyCallsSuperFinalize &  2 &  1 &  0.000 &  1.000 \\
AUTOMATE\_TESTS & EmptyStatementBlock &  2 &160 &  0.000 &  0.748 \\
WRITE\_SHORT\_UNITS & EmptyStatementBlock &  2 &160 &  0.000 &  0.748 \\
WRITE\_SIMPLE\_UNITS & EmptyStatementBlock &  4 &160 &  0.001 &  0.748 \\
\bottomrule

\toprule
\textbf{CheckStyle} &  \textbf{CoverityScan} &  \textbf{\# CHS pairs} &  \textbf{\# CS pairs} & \textbf{CHS Agr.} & \textbf{CS Agr.} \\
\midrule
FinalParametersCheck &Unsafe reflection &313 & 2 &  0.001 & 1.000 \\
InvalidJavadocPositionCheck &  IP: Ignored parameter &  4 & 2 &  0.000 & 1.000 \\
NoWhitespaceAfterCheck &  IP: Ignored parameter &  1 & 2 &  0.000 & 1.000 \\
MissingJavadocMethodCheck &  IP: Ignored parameter &112 & 2 &  0.001 & 1.000 \\
MagicNumberCheck &  IP: Ignored parameter & 50 & 2 &  0.000 & 1.000 \\
LineLengthCheck &  IP: Ignored parameter &  6 & 2 &  0.000 & 1.000 \\
JavadocVariableCheck &  IP: Ignored parameter & 36 & 2 &  0.000 & 1.000 \\
JavadocStyleCheck &  IP: Ignored parameter & 84 & 2 &  0.001 & 1.000 \\
JavadocMethodCheck &  IP: Ignored parameter & 66 & 2 &  0.001 & 1.000 \\
IndentationCheck &  IP: Ignored parameter &  1152 & 2 &  0.000 & 1.000 \\
\bottomrule

\toprule
\textbf{CheckStyle} & \textbf{FindBugs} &  \textbf{\# CHS pairs} &  \textbf{\# FB pairs} & \textbf{CHS Agr.} &  \textbf{FB Agr.} \\
\midrule
AvoidStarImportCheck &  SA\_FIELD\_SELF\_COMPUTATION &  8 & 2 &  0.000 & 1.000 \\
FinalParametersCheck & NP\_NONNULL\_FIELD\_NOT\_INITIALIZED\_IN\_CONSTRUCTOR &  3 & 1 &  0.000 & 1.000 \\
RedundantModifierCheck &  IC\_SUPERCLASS\_USES\_SUBCLASS\_DURING\_INITIALIZATION &  1 & 1 &  0.000 & 1.000 \\
DesignForExtensionCheck & RV\_RETURN\_VALUE\_OF\_PUTIFABSENT\_IGNORED &  8 & 1 &  0.000 & 1.000 \\
NeedBracesCheck &  TQ\_EXPLICIT\_UNKNOWN\_SOURCE\_VALUE\_REACHES\_NEVER... & 28 & 1 &  0.000 & 1.000 \\
JavadocVariableCheck & RV\_RETURN\_VALUE\_OF\_PUTIFABSENT\_IGNORED &  8 & 1 &  0.000 & 1.000 \\
JavadocVariableCheck &  IC\_SUPERCLASS\_USES\_SUBCLASS\_DURING\_INITIALIZATION &  1 & 1 &  0.000 & 1.000 \\
WhitespaceAroundCheck &  BIT\_IOR & 10 & 1 &  0.000 & 1.000 \\
JavadocPackageCheck &  IC\_SUPERCLASS\_USES\_SUBCLASS\_DURING\_INITIALIZATION &  1 & 1 &  0.000 & 1.000 \\
MissingJavadocMethodCheck &FI\_MISSING\_SUPER\_CALL &149 & 1 &  0.001 & 1.000 \\
\bottomrule

\end{tabular}
\end{table}

\newpage
\begin{table}[H]
 \caption{\textbf{RQ$_2$.} The 10 most recurrent rule pairs detected in the same class by the considered SATs and their corresponding agreement values.}
 \label{tab:AllToghether2}
 \centering
 \tiny
 \begin{tabular}{p{4cm}| p{4.5cm}|p{0.45cm}|p{0.45cm}|p{0.45cm}|p{0.45cm}}
 
 \toprule
\textbf{CheckStyle} &  \textbf{PMD} &  \textbf{\# CHS pairs} & \textbf{\# PMD pairs} & \textbf{CHS Agr.} &  \textbf{PMD Agr.} \\
\midrule
FinalParametersCheck &  AvoidMultipleUnaryOperators & 51 &  2 &  0.000 &  1.000 \\
RegexpSinglelineCheck &  AvoidMultipleUnaryOperators & 75 &  2 &  0.000 &  1.000 \\
NoFinalizerCheck &  FinalizeOnlyCallsSuperFinalize &  1 &  1 &  0.015 &  1.000 \\
VisibilityModifierCheck &InvalidLogMessageFormat & 15 &  1 &  0.000 &  1.000 \\
AbbreviationAsWordInNameCheck &  FinalizeOnlyCallsSuperFinalize &  1 &  1 &  0.000 &  1.000 \\
NoWhitespaceAfterCheck &  AvoidMultipleUnaryOperators &  7 &  2 &  0.000 &  1.000 \\
VariableDeclarationUsageDistanceCheck &  AvoidMultipleUnaryOperators &  7 &  2 &  0.001 &  1.000 \\
NeedBracesCheck &  AvoidMultipleUnaryOperators & 90 &  2 &  0.000 &  1.000 \\
JavadocParagraphCheck &  FinalizeOnlyCallsSuperFinalize &  1 &  1 &  0.000 &  1.000 \\
NonEmptyAtclauseDescriptionCheck &  FinalizeOnlyCallsSuperFinalize & 10 &  1 &  0.000 &  1.000 \\
\bottomrule
 
\toprule
\textbf{CoverityScan} & \textbf{FindBugs} &  \textbf{\#CS pairs} & \textbf{\#FB pairs} & \textbf{CS Agr.} &  \textbf{FB Agr.} \\
\midrule
UCF: Useless control flow & BC\_UNCONFIRMED\_CAST\_OF\_RETURN\_VALUE & 1 & 2 & 1.000 & 0.001 \\
Dead default in switch & DLS\_DEAD\_LOCAL\_STORE & 1 & 5 & 1.000 & 0.004 \\
DLS: Dead local store &NP\_ALWAYS\_NULL\_EXCEPTION & 3 & 1 & 0.010 & 1.000 \\
UCF: Useless control flow &  BC\_UNCONFIRMED\_CAST & 1 & 1 & 1.000 & 0.000 \\
Unsafe reflection &NP\_LOAD\_OF\_KNOWN\_NULL\_VALUE & 2 & 2 & 1.000 & 0.012 \\
UCF: Useless control flow &UCF\_USELESS\_CONTROL\_FLOW & 1 & 1 & 1.000 & 0.011 \\
OGNL injection &  RCN\_REDUNDANT\_NULLCHECK\_OF\_NONNULL\_VALUE & 1 & 1 & 1.000 & 0.001 \\
Failure to call super.finalize() &  FI\_NULLIFY\_SUPER & 1 & 1 & 0.500 & 1.000 \\
REC: RuntimeException capture & RV\_RETURN\_VALUE\_OF\_PUTIFABSENT\_IGNORED & 1 & 1 & 0.013 & 1.000 \\
USELESS\_STRING: Useless/non-informative string... &FI\_MISSING\_SUPER\_CALL & 5 & 1 & 0.250 & 1.000 \\
\bottomrule

\toprule
\textbf{CoverityScan} & \textbf{PMD} &  \textbf{\# CS pairs} &  \textbf{\# PMD pairs} &  \textbf{CS Agr.} &  \textbf{PMD Agr.} \\
\midrule
Dead default in switch &  AvoidInstantiatingObjectsInLoops & 1 &  1 & 1.000 &  0.000 \\
TLW: Wait with two locks held &ShortVariable & 1 & 10 & 1.000 &  0.000 \\
TLW: Wait with two locks held &  UseCorrectExceptionLogging & 1 &  3 & 1.000 &  0.010 \\
TLW: Wait with two locks held &  UseConcurrentHashMap & 1 &  5 & 1.000 &  0.002 \\
TLW: Wait with two locks held &UnusedImports & 1 & 21 & 1.000 &  0.000 \\
TLW: Wait with two locks held &  UnnecessaryFullyQualifiedName & 1 &  2 & 1.000 &  0.000 \\
TLW: Wait with two locks held &  TooManyMethods & 1 &  1 & 1.000 &  0.000 \\
TLW: Wait with two locks held &TooManyFields & 1 &  1 & 1.000 &  0.001 \\
TLW: Wait with two locks held &  StdCyclomaticComplexity & 1 &  2 & 1.000 &  0.000 \\
TLW: Wait with two locks held & ProperLogger & 1 &  2 & 1.000 &  0.001 \\
\bottomrule

\toprule
\textbf{FindBugs} & \textbf{PMD} &  \textbf{\# FB pairs} &  \textbf{\# PMD pairs} &  \textbf{FB Agr.} &  \textbf{PMD Agr.} \\
\midrule
DMI\_EMPTY\_DB\_PASSWORD &LawOfDemeter & 1 &  9 & 1.000 &  0.000 \\
TESTING &  OnlyOneReturn & 1 &169 & 1.000 &  0.002 \\
ICAST\_BAD\_SHIFT\_AMOUNT & DataflowAnomalyAnalysis & 8 &100 & 1.000 &  0.001 \\
BIT\_IOR &  SignatureDeclareThrowsException & 1 & 14 & 1.000 &  0.000 \\
BC\_IMPOSSIBLE\_DOWNCAST\_OF\_TOARRAY &  ForLoopCanBeForeach & 1 &  6 & 1.000 &  0.000 \\
LG\_LOST\_LOGGER\_DUE\_TO\_WEAK\_REFERENCE & MethodArgumentCouldBeFinal & 1 & 62 & 1.000 &  0.000 \\
QF\_QUESTIONABLE\_FOR\_LOOP & MethodArgumentCouldBeFinal & 1 &  6 & 1.000 &  0.000 \\
HRS\_REQUEST\_PARAMETER\_TO\_HTTP\_HEADER & MethodArgumentCouldBeFinal & 1 &  6 & 1.000 &  0.000 \\
ICAST\_BAD\_SHIFT\_AMOUNT &  ConfusingTernary & 8 &  4 & 1.000 &  0.000 \\
BIT\_IOR &  ConfusingTernary & 1 &  3 & 1.000 &  0.000 \\
\bottomrule

\toprule
\textbf{SonarQube} & \textbf{BetterCodeHub} &  \textbf{\# SQ pairs} &  \textbf{\# BCH pairs} &  \textbf{SQ Agr.} &  \textbf{BCH Agr.} \\
\midrule
S2275 &  COUPLE\_ARCHITECTURE\_COMPONENTS\_LOOSELY & 1 &  2 & 1.000 &  0.001 \\
S2275 &  AUTOMATE\_TESTS & 1 &  2 & 1.000 &  0.000 \\
S888 &  COUPLE\_ARCHITECTURE\_COMPONENTS\_LOOSELY & 2 &  4 & 0.667 &  0.001 \\
S888 &WRITE\_CLEAN\_CODE & 2 & 10 & 0.667 &  0.001 \\
S888 & WRITE\_CODE\_ONCE & 2 & 26 & 0.667 &  0.002 \\
S888 &  AUTOMATE\_TESTS & 2 &  2 & 0.667 &  0.000 \\
S888 &SEPARATE\_CONCERNS\_IN\_MODULES & 2 &  2 & 0.667 &  0.000 \\
ObjectFinalizeOverridenCalls SuperFinalizeCheck &  AUTOMATE\_TESTS & 1 &  3 & 0.500 &  0.000 \\
S2232 &SEPARATE\_CONCERNS\_IN\_MODULES & 1 &  1 & 0.500 &  0.000 \\
S2232 &  AUTOMATE\_TESTS & 1 &  1 & 0.500 &  0.000 \\
\bottomrule

\toprule
\textbf{SonarQube} & \textbf{CheckStyle} & \textbf{\# SQ pairs} &  \textbf{\# CHS pairs} & \textbf{ SQ Agr.} & \textbf{ CHS Agr.} \\
\midrule
S2252 &  CommentsIndentationCheck & 1 &  1 & 1.000 &  0.000 \\
S2200 &  NeedBracesCheck & 2 & 12 & 1.000 &  0.000 \\
S2123 & RedundantModifierCheck & 2 & 56 & 1.000 &  0.002 \\
S2123 &  InvalidJavadocPositionCheck & 2 &  4 & 1.000 &  0.000 \\
S2123 &  RegexpSinglelineCheck & 2 & 42 & 1.000 &  0.000 \\
S2123 &  WhitespaceAroundCheck & 2 & 22 & 1.000 &  0.000 \\
S2200 & EqualsHashCodeCheck & 2 &  1 & 1.000 &  0.002 \\
S2200 &  FileTabCharacterCheck & 2 &  8 & 1.000 &  0.000 \\
S2200 &FinalParametersCheck & 2 &  9 & 1.000 &  0.000 \\
S2200 & IndentationCheck & 2 &  1 & 1.000 &  0.000 \\
\bottomrule

\toprule
\textbf{SonarQube} & \textbf{CoverityScan} & \textbf{ \# SQ pairs} &  \textbf{\# CS pairs} &  \textbf{SQ Agr.} & \textbf{ CS Agr.} \\
\midrule
S1244 &  Unexpected control flow & 2 & 2 & 0.001 & 1.000 \\
S00105 &  Use of hard-coded cryptographic key & 1 & 1 & 0.000 & 1.000 \\
S1125 &Dead default in switch & 4 & 1 & 0.001 & 1.000 \\
S1126 &Dead default in switch & 2 & 1 & 0.001 & 1.000 \\
S1132 &Dead default in switch & 1 & 1 & 0.000 & 1.000 \\
S1149 &Dead default in switch & 1 & 1 & 0.000 & 1.000 \\
S1151 &Dead default in switch & 8 & 1 & 0.001 & 1.000 \\
S1172 &Dead default in switch & 4 & 1 & 0.002 & 1.000 \\
S1213 &Dead default in switch &26 & 1 & 0.001 & 1.000 \\
S1226 &Dead default in switch & 7 & 1 & 0.001 & 1.000 \\
\bottomrule

\end{tabular}
\end{table}

\end{document}